\documentclass[pra,twocolumn,preprintnumbers,amsmath,amssymb,floatfix,longbibliography,superscriptaddress]{revtex4-2}
\usepackage{graphicx}
\usepackage{graphics}
\usepackage{psfrag}
\usepackage{hyperref}
\usepackage{multirow}
\usepackage[sort&compress]{natbib}
\usepackage{amssymb}
\usepackage{amsmath}
\usepackage{amsthm}
\usepackage{bbold}
\usepackage{bbm}
\usepackage{enumerate}
\usepackage{textcomp}

\newcommand{\vare}{\varepsilon}

\newcommand{\rmi}{{\rm i}}

\DeclareMathOperator{\artanh}{artanh}

\usepackage{xcolor}

\begin{document}

\hypersetup{pdftitle={title}}
\title{Crystallography of Hyperbolic Lattices}

\author{Igor Boettcher}
\email{iboettch@ualberta.ca}
\affiliation{Department of Physics, University of Alberta, Edmonton, Alberta T6G 2E1, Canada}
\affiliation{Theoretical Physics Institute, University of Alberta, Edmonton, Alberta T6G 2E1, Canada}

\author{Alexey V. Gorshkov}
\affiliation{Joint Quantum Institute, NIST/University of Maryland, College Park, MD 20742, USA}
\affiliation{Joint Center for Quantum Information and Computer Science, NIST/University of Maryland, College Park, Maryland 20742, USA}

\author{Alicia J. Koll\'{a}r}
\affiliation{Joint Quantum Institute, NIST/University of Maryland, College Park, MD 20742, USA}

\author{Joseph Maciejko}
\affiliation{Department of Physics, University of Alberta, Edmonton, Alberta T6G 2E1, Canada}
\affiliation{Theoretical Physics Institute, University of Alberta, Edmonton, Alberta T6G 2E1, Canada}

\author{Steven Rayan}
\affiliation{Department of Mathematics and Statistics, University of Saskatchewan, Saskatoon, Saskatchewan S7N 5E6, Canada}
\affiliation{Centre for Quantum Topology and Its Applications (quanTA), University of Saskatchewan, Saskatoon, Saskatchewan S7N 5E6, Canada}

\author{Ronny Thomale}
\affiliation{Institute for Theoretical Physics and Astrophysics, University of W\"{u}rzburg, Am Hubland, D-97074 W\"{u}rzburg, Germany}

\begin{abstract}
Hyperbolic lattices are a revolutionary platform for tabletop simulations of holography and quantum physics in curved space and facilitate efficient quantum error correcting codes. Their underlying geometry is non-Euclidean, and the absence of Bloch's theorem precludes the straightforward application of the often indispensable energy band theory to study model Hamiltonians on hyperbolic lattices. Motivated by recent insights into hyperbolic band theory, we initiate a crystallography of hyperbolic lattices. We show that many hyperbolic lattices feature a hidden crystal structure characterized by unit cells, hyperbolic Bravais lattices, and associated symmetry groups. Using the mathematical framework of higher-genus Riemann surfaces and Fuchsian groups, we derive, for the first time, a list of example hyperbolic $\{p,q\}$ lattices and their hyperbolic Bravais lattices, including five infinite families and several graphs relevant for experiments in circuit quantum electrodynamics and topolectrical circuits. This dramatically simplifies the computation of energy spectra of tight-binding Hamiltonians on hyperbolic lattices, from exact diagonalization on the graph to solving a finite set of equations in terms of irreducible representations. The significance of this achievement needs to be compared to the all-important role played by conventional Euclidean crystallography in the study of solids. We exemplify the high potential of this approach by constructing and diagonalizing finite-dimensional Bloch wave Hamiltonians. Our work lays the foundation for generalizing some of the most powerful concepts of solid state physics, crystal momentum and Brillouin zone, to the emerging field of hyperbolic lattices and tabletop simulations of gravitational theories, and reveals the connections to concepts from topology and algebraic geometry.
\end{abstract}

\maketitle

Hyperbolic geometry plays a paramount role at the frontier of both theoretical and experimental physics. It underlies holographic descriptions of strongly coupled systems and models for quantum chaos, quantum gravity, and quantum entanglement \cite{BALAZS1986109,maldacena1999large,witten1998anti,PhysRevLett.96.181602}. It is fundamental to modern computational many-body techniques \cite{PhysRevLett.99.220405,PhysRevLett.101.110501,PhysRevD.86.065007,swingle2012constructing,PhysRevLett.110.100402,Bao_2017,milsted2018geometric} and forms the basis for powerful quantum error correcting codes \cite{pastawski2015holographic,Breuckmann_2016,Breuckmann_2017,Lavasani_2019,jahn2021holographic}. Recent experimental realizations of hyperbolic lattices \cite{kollar2019hyperbolic,kollar2019line,lenggenhager2021electric} in circuit quantum electrodynamics (QED) \cite{houck2012chip,PhysRevA.86.023837,schmidt2013circuit,PhysRevX.6.041043,GU20171} and topolectrical circuits \cite{PhysRevX.5.021031,PhysRevLett.114.173902,lee2018topolectrical,imhof2018topolectrical,PhysRevB.99.161114,stegmaier2020topological} have set the stage for the quantum simulation of curved space physics using discrete geometries \cite{PhysRevX.10.011009,PhysRevLett.125.053901,PhysRevA.102.032208,brower2019lattice,kollar2021gap,PhysRevD.102.034511,maciejko2020hyperbolic,zhang2020efimov,PhysRevA.103.033703,zhu2021quantum,ikeda2021,bienias2021circuit}. The corresponding non-Euclidean graphs are suited to be implemented in various other topological photonics platforms \cite{RevModPhys.91.015006}. These cutting-edge efforts complement important previous experimental simulations of curved space using optical metamaterials \cite{Philbin_2008,PhysRevLett.84.822,genov2009mimicking,PhysRevLett.105.067402,Chen10,bekenstein2015optical,bekenstein2017control}, ultracold quantum gases \cite{PhysRevLett.85.4643,PhysRevLett.91.240407,PhysRevA.70.063615,carusotto2008numerical,PhysRevLett.105.240401,Boada_2011,PhysRevD.85.044046,Steinhauer_2016,PhysRevLett.118.130404,Kosior_2018,Hu_2019}, electromagnetic waveguides \cite{PhysRevLett.95.031301}, trapped ions \cite{PhysRevLett.104.250403}, and other platforms \cite{PhysRevLett.106.021302,Sheng_2013}.

Stimulated by these intimate connections to outstanding open problems in physics, strong interest in the properties of hyperbolic space resurged in the last two decades. However, while many important results have been obtained since the 19$^{\rm th}$ century \cite{BookMagnus,BookCoxeter,BookNumber,Cannon,BookBujalance}, several critical questions about hyperbolic space that are relevant to physicists remain unanswered. Most strikingly, perhaps, due to the absence of Bloch's theorem, the energy spectrum of a single particle hopping on a hyperbolic lattice can only be obtained by exact numerical diagonalization of the Hamiltonian. This preempts any treatment of macroscopically large systems even in the noninteracting limit. One way around this issue is to concentrate on long-wavelength excitations, where a continuum approximation can capture several features of the discrete spectrum \cite{PhysRevA.102.032208}. To also resolve excitations with higher energies, an alternative approach is to study Bloch waves on hyperbolic tessellations, which leads to the hyperbolic band theory of Ref. \cite{maciejko2020hyperbolic}. In order to develop a complete band structure theory for hyperbolic lattices that can capture all single-particle eigenstates, on the other hand, it is mandatory to first identify the crystallographic symmetries of the lattice and then construct wavefunctions from their representations \cite{BookDresselhaus}. In this work, we address the first part of the problem and outline a Bloch wave theory to answer aspects of the second part.

\begin{figure}[t]
\centering
\includegraphics[width=8.5cm]{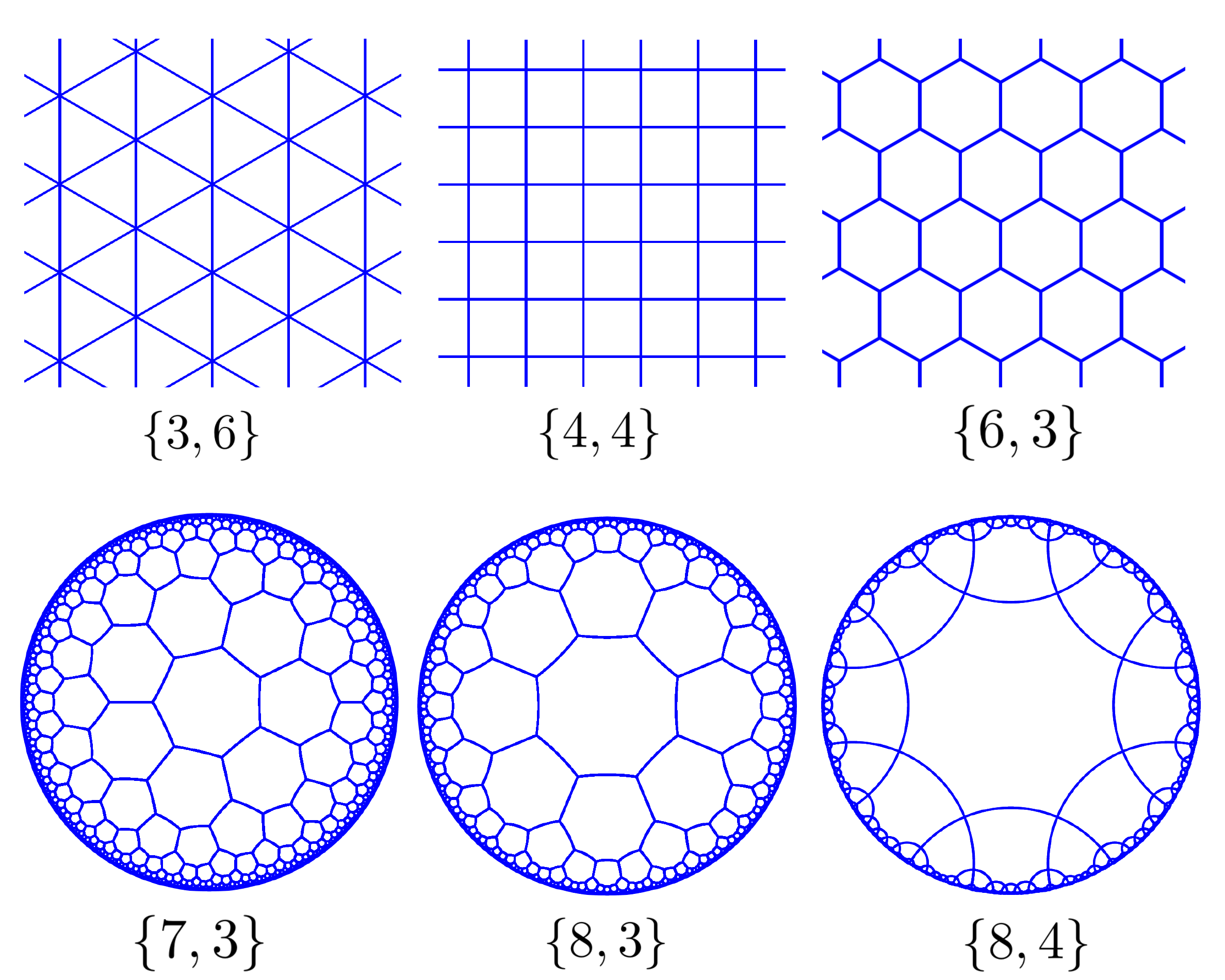}
\caption{We consider $\{p,q\}$ lattices, which are made from regular $p$-gons such that the coordination number of each lattice site is $q$. The well-known triangular ($\{3,6\}$), square ($\{4,4\}$), and hexagonal ($\{6,3\}$) lattices (first row) constitute tessellations of the Euclidean plane. Hyperbolic lattices, defined by $ (p-2)(q-2)>4$, are tessellations of the hyperbolic plane of constant negative curvature. We show three examples (second row), with hyperbolic space represented by the Poincar\'{e} disk model, reviewed in Sec. \ref{SecHyp}, where the non-Euclidean metric is such that the distance between any two neighboring sites in a hyperbolic lattice is equal. Lattice sites are connected by geodesic lines, which are circular arcs that (when extended) intersect the disk boundary orthogonally. The circular nature of geodesics in the $\{7,3\}$ and $\{8,3\}$ lattices is less easily visible compared to the $\{8,4\}$ lattice, but nonetheless present.}
\label{FigIntro}
\end{figure}

The hyperbolic lattices we consider are of $\{p,q\}$ type, which means that they are tessellations of the plane by regular $p$--gons such that each lattice site has coordination number $q$. For $(p-2)(q-2)=4$, such lattices are tilings of the Euclidean plane by regular polygons, which can only be achieved by triangles, squares, and hexagons, corresponding to the solutions $\{3,6\}$, $\{4,4\}$, and $\{6,3\}$, see Fig. \ref{FigIntro}. On the other hand, for $(p-2)(q-2)>4$, we obtain a tessellation of the hyperbolic plane which we call a \emph{hyperbolic lattice}. Some examples are shown in Fig. \ref{FigIntro}. Obviously, there are infinitely many integer solutions $p$ and $q$ to this inequality, implying striking lattice properties. For instance, hyperbolic lattices can have any $p$--fold rotation symmetry, in stark contrast to Euclidean lattices. The high connectivity of such lattices also implies that their number of sites grows exponentially in the graph diameter (which is the shortest number of steps to get from one end of the lattice to the other). To embed these lattices into hyperbolic space, i.e.~assign a complex coordinate to each lattice site, we use the Poincar\'{e} disk model, which is reviewed in Sec. \ref{SecHyp}.

The immense value of crystallography in the theory of solids stems from the ability to utilize crystal symmetries to divide macroscopic numbers of lattice sites into unit cells that are arranged in a well-known manner in a Bravais lattice. In fact, in a bottom-up approach, we may construct every two-dimensional Euclidean lattice from a finite set of points $\{z^{(1)},\dots,z^{(N)}\}$, called the unit cell, which is repeated periodically in a Bravais lattice specified by two primitive translation vectors $\textbf{a}_1$ and $\textbf{a}_2$. (We are restricting ourselves to symmorphic space groups here for simplicity.) Every lattice site is then uniquely defined by a pair of numbers $(a,\textbf{n})$, where $a\in\{1,\dots,N\}$ is the position inside the unit cell, and $\textbf{n}=(n_1,n_2)\in\mathbb{Z}^2$ locates the unit cell within the Bravais lattice at position $n_1\textbf{a}_1+n_2\textbf{a}_2$. An analogous construction applies to Euclidean lattices in higher dimensions, and it is a famous result that the number of distinct Euclidean Bravais lattices is finite in every dimension. In two and three dimensions, there are 5 and 14 Bravais lattices, respectively.

\begin{figure}[t]
\centering
\includegraphics[width=8.5cm]{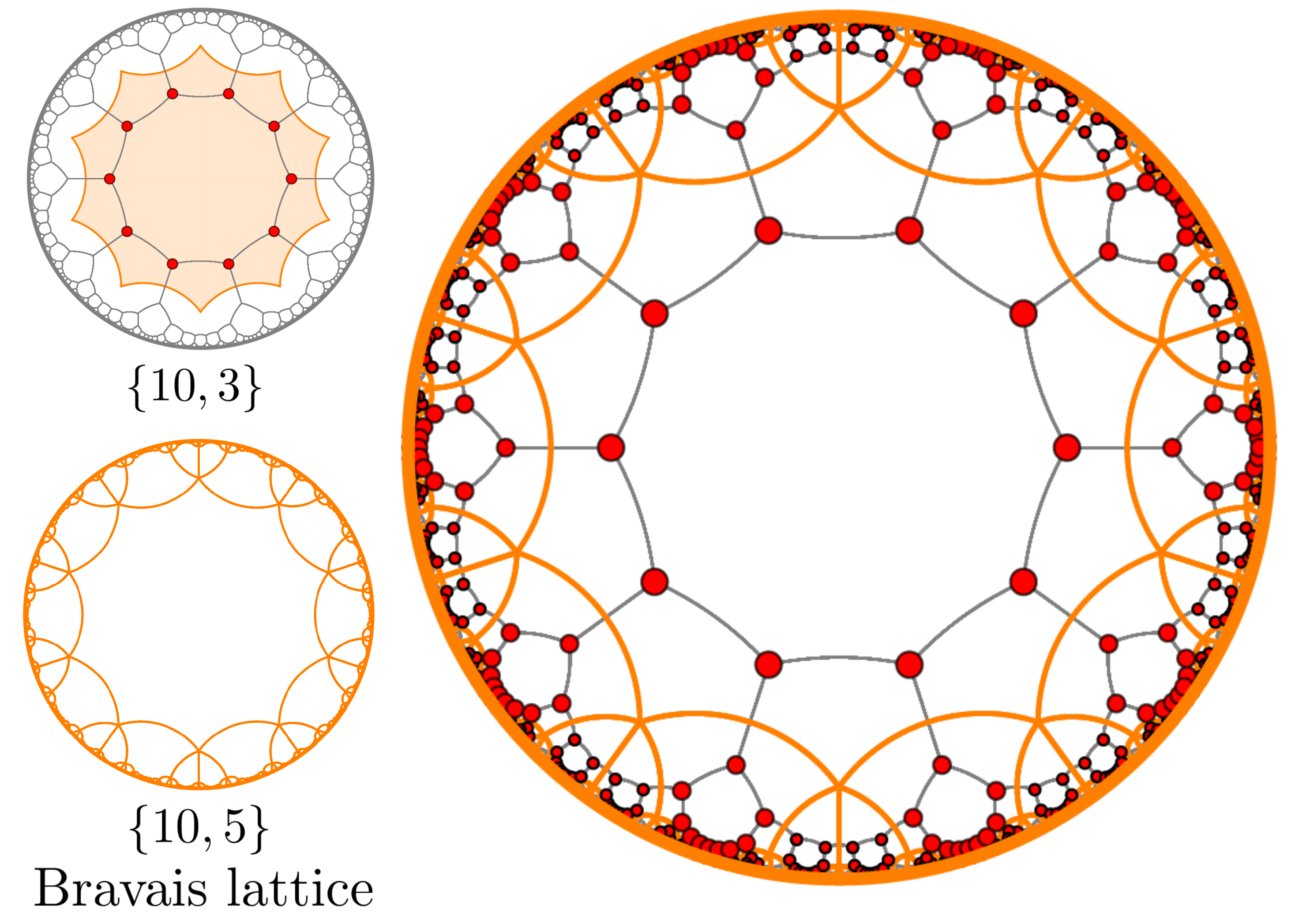}
\caption{In this work, we develop the formalism how to apply the crystallographic notions of unit cell and Bravais lattice to hyperbolic lattices. Here we give an example. \emph{Top left.} The unit cell of the $\{10,3\}$ lattice consists of ten lattice sites (marked red). The associated analogue of the Wigner--Seitz cell or fundamental domain is the decagon of the $\{10,5\}$ lattice (shown in orange). \emph{Bottom left.} Therefore, the $\{10,5\}$ lattice constitutes the Bravais lattice of the $\{10,3\}$ lattice. \emph{Right.} By filling the decagons of the $\{10,5\}$ lattice with the ten sites of the unit cell, we obtain all sites of the $\{10,3\}$ lattice.\\}
\label{Fig103and105}
\end{figure}

Once the unit cell and Bravais lattice of a given Euclidean lattice are identified, the single-particle eigenstates can be constructed from representations of the translation operators $\hat{T}_{\textbf{n}}$ of the Bravais lattice, mapping the origin to $n_1\textbf{a}_1+n_2\textbf{a}_2$. Due to Bloch's theorem, all of these representations are one-dimensional and labeled by crystal momenta $\textbf{q}=(q_1,q_2)$, which have as many components as there are primitive translation vectors, yielding $N$ energy bands. It is important to realize that Bloch's theorem, although convenient in the construction of the eigenstates, is not a necessary piece. If the translations $\hat{T}_{\textbf{n}}$ were not mutually commuting, then their representations would not all be one-dimensional. Still we could find these representations, label them by certain quantum numbers, and use this structure to construct $N$ energy bands. A well-known example is, of course, the eigenstates of a particle in a three-dimensional spherically symmetric potential, where states are labeled by the usual quantum numbers $(n,\ell,m)$, with the dimension of each eigenspace or representation being $2\ell+1$, since three-dimensional rotations generally do not commute.

In order to solve the spectral problem for hyperbolic lattices, it is therefore crucial to first generalize the concepts of unit cell and Bravais lattice to hyperbolic lattices. Only as the second step we need to worry about the representations of the generators of the Bravais lattice. It may not be obvious that the first step can be taken at all. The central result of this work is to provide, for the first time, a comprehensive list of infinitely many and experimentally relevant examples of $\{p,q\}$ lattices with their unit cells and corresponding Bravais lattices. In Fig. \ref{Fig103and105} we illuminate the example of the $\{10,3\}$ lattice, with a unit cell of ten sites, whose Bravais lattice is the $\{10,5\}$ lattice. As expected, the generators of the hyperbolic Bravais lattice, which are constructed explicitly in Sec. \ref{SecFuchs}, do not commute. More examples of $\{p,q\}$ lattices and their unit cells and Bravais lattices are collected in Tables \ref{TabUnits2} and \ref{TabUnits}.

\begin{figure}[t]
\centering
\includegraphics[width=8.5cm]{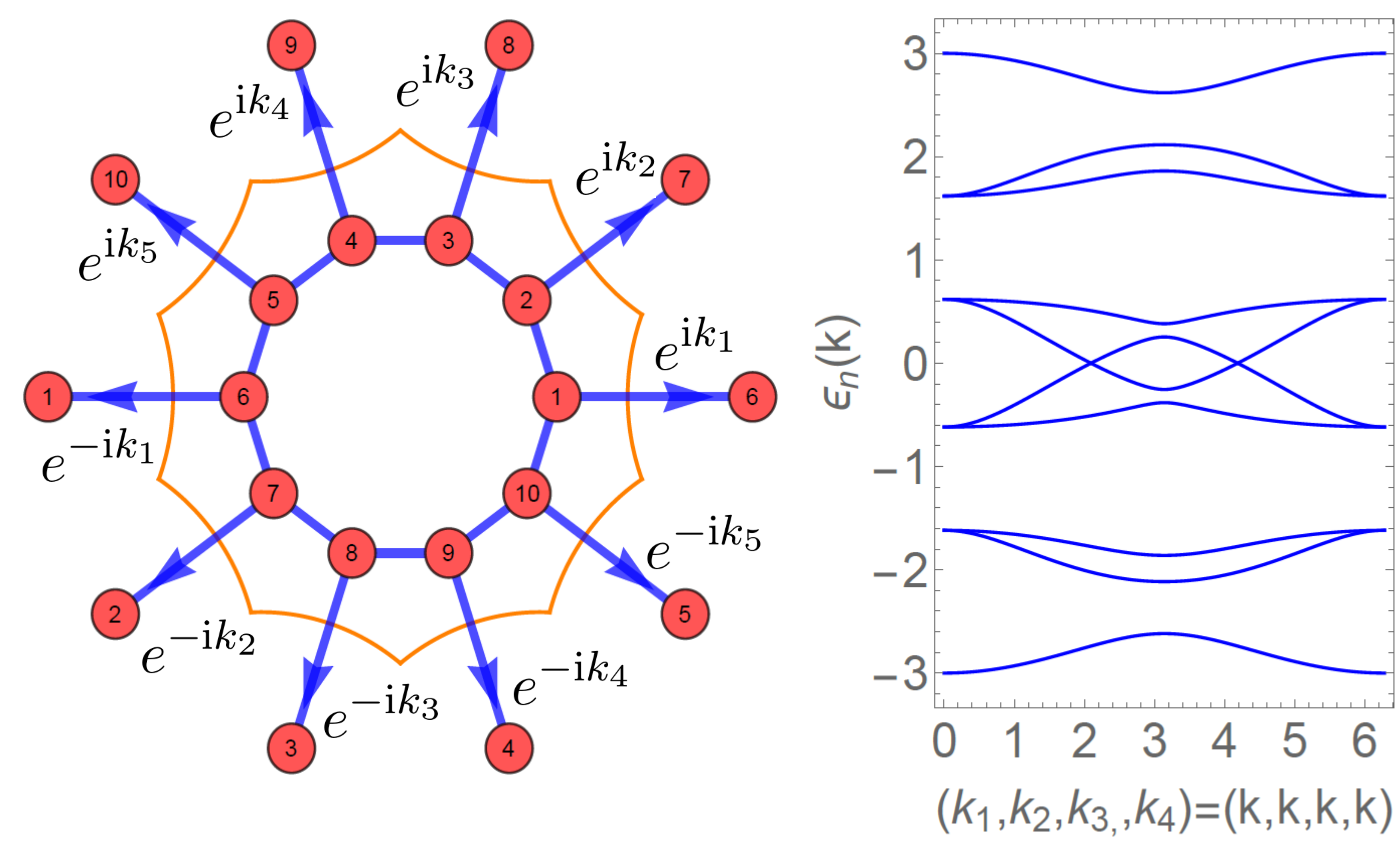}
\caption{In hyperbolic band theory, Bloch waves pick up a phase factor $e^{\rmi k_\mu}$ when going from one unit cell to the other, or, equivalently, when traversing the boundary of the fundamental domain. \emph{Left.} Continuing the example of the $\{10,3\}$ lattice from Fig. \ref{Fig103and105}, we endow the unit cell with periodic boundary conditions and obtain a graph with ten sites and coordination number 3 (red circles). The ten edges of the fundamental domain (orange) give rise to five naive momentum components, $(k_1, k_2, k_3, k_4, k_5)$, only four of which are independent. (A similar redundancy occurs in the Euclidean hexagonal lattice, see Sec. \ref{SecEucl}.) \emph{Right.} We show the band structure of the associated Bloch wave Hamiltonian in \emph{four-dimensional momentum space} along the $\textbf{k}=(k_1,k_2,k_3,k_4)^T=k(1,1,1,1)^T$ direction as a function of $k$. Strikingly, position space and momentum space do not have the same dimension for hyperbolic Bloch waves.}
\label{FigBands}
\end{figure}

Finally, let us have a glimpse at the Bloch wave band structure that is implied by the crystallography presented in this work. The tight-binding Hamiltonian we would like to diagonalize is
\begin{align}
 \label{intro1} \hat{H} = -\sum_{i,j} A_{ij} \hat{a}^\dagger_i \hat{a}_j,
\end{align}
where the sum runs over the sites of the lattice, $\hat{a}^\dagger_i$ is the creation operator of a particle at site $i$, and $A_{ij}$ is the adjacency matrix of the hyperbolic lattice. ($A$ is the matrix with entry $1$ if $i$ and $j$ are connected by an edge, and zero otherwise.) Following the idea of Ref. \cite{maciejko2020hyperbolic}, we further assume that some eigenstates transform as one-dimensional representations under the generators of the Bravais lattice and, therefore, are simply Bloch waves. Roughly speaking, when going from one unit cell to the other, Bloch waves  $\psi_{\textbf{k}}(z_i)$ pick up a phase factor $e^{\rmi k_\mu}$, see Fig. \ref{FigBands} for an illustration on the example of the $\{10,3\}$ lattice discussed earlier. The number of independent momentum components of such a Bloch wave is $2g$, where $g\geq 2$ is the unique genus of a Riemann surface that can be covered by the fundamental domain, see Sec. \ref{SecCrys}. Projecting the operator $\hat{a}_i$ to the space spanned by Bloch waves, we arrive at the problem of diagonalizing the Bloch wave Hamiltonian
\begin{align}
  \label{intro2} \hat{H}_{\rm BW} = -\sum_{\textbf{k}}  \sum_{a,a'=1}^N\bar{A}_{aa'}(\textbf{k}) \hat{a}^\dagger_{\textbf{k}a}\hat{a}_{\textbf{k}a'},
\end{align}
with $\bar{A}(\textbf{k})$ the $N\times N$ adjacency matrix of the unit cell endowed with periodic boundary conditions, whose edges are labeled by entries $1$ and $e^{\rmi k_{\mu}}$ in a well-specified manner. The eigenvalues of the matrix $\bar{A}(\textbf{k})$ yield $N$ energy bands that constitute the Bloch wave spectrum of the given $\{p,q\}$ lattice. An example band structure is shown in Fig. \ref{FigBands}, with the corresponding matrix $\bar{A}(\textbf{k})$ given by Eq. (\ref{app8}).  The construction of Bloch wave Hamiltonians and their spectra are discussed in Sec. \ref{SecBloch}.

We emphasize that every Bloch wave is a solution to the tight-binding problem on the hyperbolic lattice and thus yields a valid eigenenergy. We merely lack the information on the fraction of eigenstates that transform under a higher-dimensional representation. Some first steps towards a complete classification of irreducible representations and hence a Bloch theorem for hyperbolic Bravais lattices have been taken in Ref. \cite{maciejko2021automorphic} for finite patterns of $\{8,8\}$ type. One remarkable finding is that for many choices of such finite patterns, all irreducible representations are one-dimensional and hence Bloch wave theory is exact. On the other hand, instances where two-dimensional representations play a role could also be identified.

The main result of this paper is to provide a concrete list of example $\{p,q\}$ lattices with their unit cells and Bravais lattices, including several cases for genus $g=2,3$ and five infinite families. To the best of our knowledge, no such list has been collected before, and we are convinced that it will fundamentally alter the perspective on hyperbolic lattices in future theoretical and experimental work. We give a comprehensive introduction to the mathematical toolbox required to work with hyperbolic crystallography in practice. We expect these concepts from topology and geometry, although covered in mathematics textbooks, to be less known to the wider physics community. In particular, our construction is strongly built on the notion of patterns on higher-genus Riemann surfaces from Ref. \cite{Edmonds}. Our identification of hyperbolic Bravais lattices parallels the study of periodic boundary conditions on the hyperbolic plane of Ref. \cite{sausset2007periodic}.

This paper is organized as follows. We first review hyperbolic geometry and the Poincar\'{e} disk model in Sec. \ref{SecHyp}. We then discuss the notion of patterns on higher-genus Riemann surfaces in Sec. \ref{SecPat}, which, in particular, allows us to identify all potential regular hyperbolic Bravais lattices. In Sec. \ref{SecCrys}, we construct the generators of hyperbolic Bravais lattices and study the associated Fuchsian groups. In Sec. \ref{SecReg}, we discuss a selection of $\{p,q\}$ lattices with their unit cells and Bravais lattices. In Sec. \ref{SecAllApps}, we apply our findings to tight-binding Hamiltonians on infinite hyperbolic lattices. In App. \ref{SecGen}, we describe a method to efficiently generate large lattices. Further technical details are collected in Apps. \ref{AppFuchs}--\ref{AppSearch} and referenced in the main text.

\section{Hyperbolic Geometry}\label{SecHyp}

In this section, we introduce the Poincar\'{e} disk model of hyperbolic space and discuss its distance preserving maps. Using hyperbolic trigonometry, we construct regular hyperbolic polygons, which are the basic building block of our crystallography.

For the study of hyperbolic lattices, we employ the Poincar\'{e} disk model of hyperbolic space \cite{BALAZS1986109,Cannon}, which consists of all points in the unit disk $\mathbb{D}=\{z\in \mathbb{C},\ |z|<1\}$ equipped with the hyperbolic metric
\begin{align}
 \label{hyp1} \mbox{d}s^2 = (2\kappa)^2\ \frac{\mbox{d}x^2+\mbox{d}y^2}{(1-|z|^2)^2}.
\end{align}
Herein, $\kappa$ is the curvature radius, which sets the relevant length scale in hyperbolic space. The corresponding constant negative curvature is $K=-\kappa^{-2}$. We denote $z=x+\rmi y = r e^{\rmi \phi}$. The hyperbolic distance between two points $z,z'\in \mathbb{D}$ is given by
\begin{align}
 \label{hyp2} d(z,z') = \kappa\ \text{arcosh}\Bigl(1+\frac{2|z-z'|^2}{(1-|z|^2)(1-|z'|^2)}\Bigr).
\end{align}
The angle between two intersecting lines in $\mathbb{D}$ is given by the usual Euclidean angle. The geodesics of the Poincar\'{e} disk model are circular arcs that intersect the boundary of $\mathbb{D}$ orthogonally. This includes straight lines through the origin.

The isometries of $\mathbb{D}$ are the maps that preserve the hyperbolic distance. They may either preserve or change orientation. The orientation preserving ones are given by fractional linear transformations
\begin{align}
 \label{hyp3} z \mapsto Mz := \frac{a z +b}{b^* z+a^*}
\end{align}
with complex numbers $a$ and $b$ satisfying $|a|^2-|b|^2=1$. Identifying $M$ with the $\text{SU}(1,1)$-matrix
\begin{align}
 \label{hyp4} M = \begin{pmatrix} a & b \\ b^* & a^* \end{pmatrix},
\end{align}
the orientation preserving-maps form the group
\begin{align}
 \label{hyp5} \mathcal{P}=\text{PSU}(1,1) = \text{SU}(1,1)/\{\pm\mathbb{1}\}.
\end{align}
In the following, we denote the unit element of $\mathcal{P}$ by $1_{\mathcal{P}}$. The statement $X=1_{\mathcal{P}}$ means $X=\pm \mathbb{1}$ in the two-dimensional representation of Eq. (\ref{hyp4}). Note that if we embedded the hyperbolic lattice into the Poincar\'{e} upper-half plane $\mathbb{H}$ instead, the group of orientation preserving isometries would be $\text{PSL}(2,\mathbb{R})$, which is  isomorphic to $\mathcal{P}$.

A typical orientation reversing map in the plane is given by complex conjugation, $z\mapsto z^*$. The group of all orientation reversing isometries of $\mathbb{D}$ is given by linear fractional transformations
\begin{align}
 \label{hyp6} z \mapsto \frac{a z^* +b}{b^* z^*+a^*},
\end{align}
and so is also isomorphic to $\mathcal{P}$. Thus, every isometry of the Poincar\'{e} disk can be uniquely decomposed into an orientation-preserving one that is either combined or not combined with the map $z\mapsto z^*$. Formally, the full isometry group of $\mathbb{D}$ is, therefore, the semi-direct product $\mathcal{P}\ltimes \mathbb{Z}_2$. An analogous, but potentially more familiar situation arises for the Euclidean orthogonal group $\text{O}(2)=\text{SO}(2)\ltimes \mathbb{Z}_2$, as every element from $\text{O}(2)$ can be uniquely written as a proper rotation from $\text{SO}(2)$ that is either combined or not combined with a reflection $(x,y)\mapsto(x,-y)$.

The central building blocks of the crystallography presented in this work are regular geodesic polygons. A polygon is called regular if its internal angles are equal and its side lengths are equal. It is called geodesic if its vertices are connected by (uniquely determined) geodesic lines. The circumradius, or simply \emph{radius} hereafter, of the polygon is the distance from the center to any of its vertices. In the Euclidean plane, the internal angles of a regular $p$-gon sum up to $(p-2)\pi$, whereas the radius can be of arbitrary size. In contrast, in the Poincar\'{e} disk, the internal angles can have any value, as long as they sum up to a number \emph{smaller} than $(p-2)\pi$, while the radius is uniquely determined by the values of the angles. Generally, smaller internal angles imply larger radii. An extreme example is a polygon with all internal angles approaching zero, so that the vertices of this polygon approach the boundary of $\mathbb{D}$. In the $\{p,q\}$ lattice, the internal angles of each $p$-gon are $2\pi/q$. Consequently, to give another useful example, a decagon in the $\{10,3\}$ lattice has smaller radius, and thus smaller area, than a decagon in the $\{10,5\}$ lattice, as can be seen in Fig. \ref{Fig103and105}.

\begin{figure}[t]
\centering
\includegraphics[width=7cm]{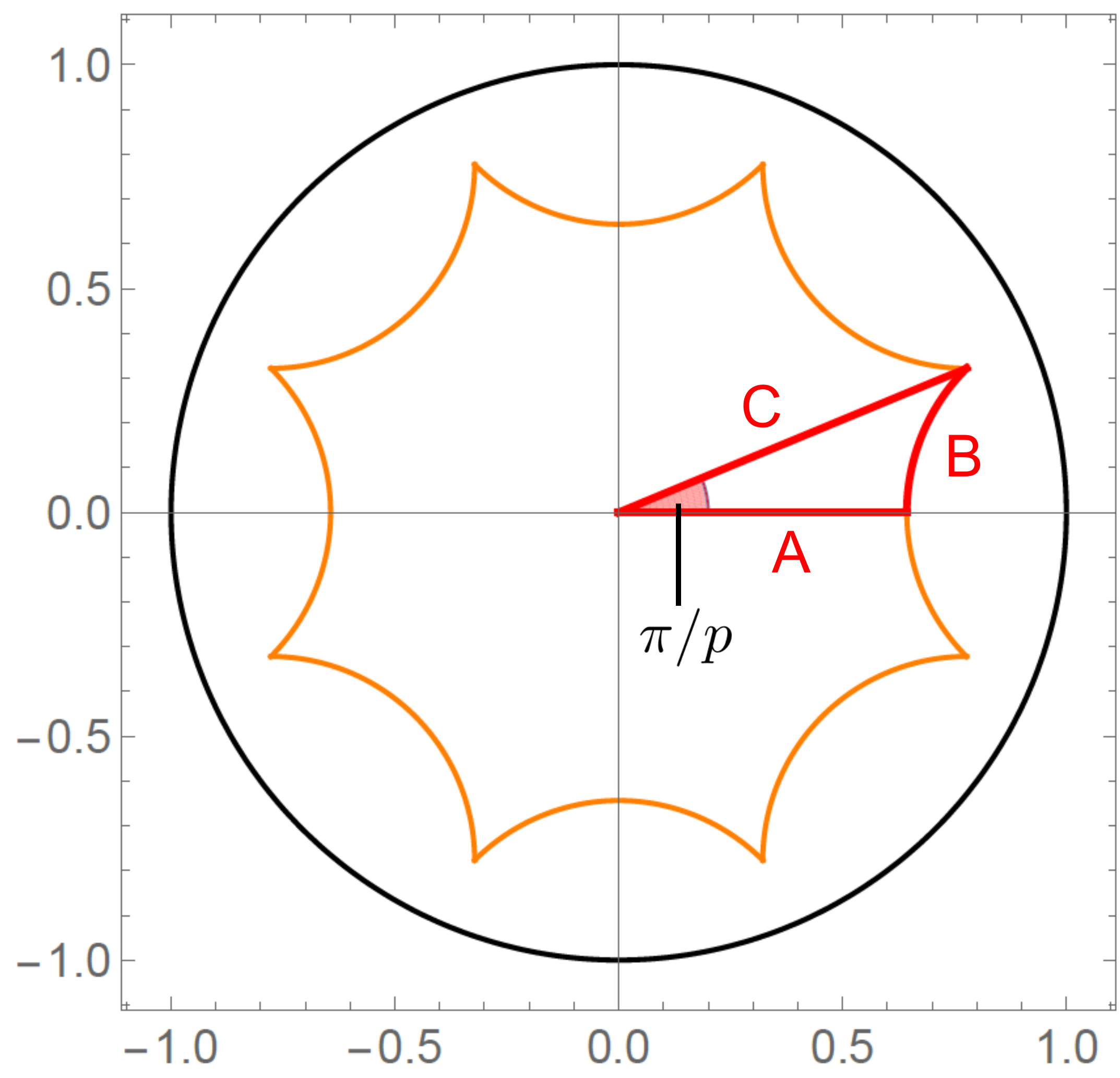}
\caption{Regular hyperbolic polygon, exemplified here by the central octagon of the $\{p,q\}=\{8,8\}$ lattice. The hyperbolic lengths $A,B,C$ enclose a right triangle with interior angles $\pi/p$ (red shaded), $\pi/2$, and $\pi/q$. They obey the relations of hyperbolic trigonometry in Eqs. (\ref{hyp8}) and (\ref{hyp9}). The Euclidean distance between the origin and the vertices is $r_0$ from Eq. (\ref{hyp7}), with $C=d(r_0,0)$. The polygon intersects the positive real axis orthogonally at $a\in \mathbb{D}$, with $a$ given by Eq. (\ref{hyp10}) and $A=d(a,0)$. The orange lines are parametrized by Eq. (\ref{hyp12}).}
\label{FigTrig}
\end{figure}

We now compute the characteristic lengths of regular $p$-gons in $\{p,q\}$ lattices. It is important to notice that we can express lengths either in terms of hyperbolic distances, given by Eq. (\ref{hyp2}) with the natural unit of length being $\kappa$, or in terms of their coordinates in the Poincar\'{e} disk, with the natural length scale given by the disk radius, which we set to unity. Denote the vertices of the polygon by $z_j = r_0 e^{\rmi  (2\pi j/p+\delta)}$, $j=1,\dots,p$, with $\delta$ an arbitrary phase. Then the radius $r_0$ of the polygon in units of the disk radius is given by
\begin{align}
\label{hyp7} r_0 = \sqrt{\frac{\cos(\frac{\pi}{p}+\frac{\pi}{q})}{\cos(\frac{\pi}{p}-\frac{\pi}{q})}}.
\end{align}
The corresponding hyperbolic radius is $C=d(r_0,0)$. Two other lengths of interest are the shortest hyperbolic distance from the center of the polygon to an edge, denoted $A$, and the hyperbolic side length, denoted $2B$, see Fig. \ref{FigTrig}. Then $A,B,C$ form a hyperbolic right triangle with internal angles  $\pi/p$, $\pi/q$, and $\pi/2$, and the rules of hyperbolic trigonometry yield the relations
\begin{align}
 \label{hyp8} \cos\Bigl(\frac{\pi}{p}\Bigr) &= \frac{\tanh(A/\kappa)}{\tanh(C/\kappa)},\\
 \label{hyp9} \sin\Bigl(\frac{\pi}{p}\Bigr) &= \frac{\sinh(B/\kappa)}{\sinh(C/\kappa)}.
\end{align}
If the polygon is oriented such that $A$ lies along the positive real axis, and accordingly $z_1 = r_0 e^{\rmi \pi/p}$ is the first vertex, then Eq. (\ref{hyp8}) implies that the edge intersects the positive real axis at a real coordinate $a\in \mathbb{D}$, with $A=d(a,0)$, determined by
\begin{align}
 \label{hyp10} \cos\Bigl(\frac{\pi}{p}\Bigr) = \frac{\tanh[2\ \text{artanh} (a)]}{\tanh[2\ \text{artanh} (r_0)]}.
\end{align}
Note that $\tanh[2\ \text{artanh}(x)]=2x/(1+x^2)$. Equations (\ref{hyp8})-(\ref{hyp10}) reproduce the Euclidean result for $a,r_0\ll 1$, because the hyperbolic metric in Eq. (\ref{hyp1}) becomes flat for $|z|\ll 1$.

We conclude this section with a parametrization of the geodesic arcs that comprise the edges of a regular hyperbolic $p$-gon.  Denote the $p$ sides by
\begin{align}
 \label{hyp12} \mathcal{C}_\mu = \Bigl\{ (c-\rho e^{\rmi \theta})e^{\rmi 2\pi (\mu-1)/p},\ \theta\in[-\theta_0,\theta_0]\ \Bigr\}
\end{align}
with $\mu=1,\dots,p$. The parameters $c,\rho,\theta_0$ are determined by $c-\rho = a$ and $c-\rho e^{-\rmi \theta_0} = z_1 = r_0 e^{\rmi \pi/p}$. This is solved by
\begin{align}
 \label{hyp13} \rho &= \frac{a^2-2ar_0\cos(\pi/p)+r_0^2}{2r_0\cos(\pi/p)-2a},\\
 \label{hyp14} c&= a+\rho,\ \sin(\theta_0) = \frac{r_0}{\rho}\sin(\pi/p).
\end{align}
The value of the internal angles ($2\pi/q$) is arbitrary and enters through $r_0$. In principle, $q$ could be noninteger-valued, although this situation does not arise in the applications considered here. The formulas derived in this section are also valid for odd values of $p$, say, for a regular hyperbolic 7-gon.

\section{Patterns}\label{SecPat}

In this section we discuss the concept of patterns, which are finite hyperbolic graphs embedded into closed Riemann surfaces, and which determine the size and shape of the unit cell of hyperbolic lattices. For this purpose we first recall the classification of Riemann surfaces $M$ via their number of holes. We then describe how to determine which patterns can be drawn onto which surfaces and identify those patterns that cover the surface with a single face, because they are particularly important for hyperbolic crystallography.

Let $M$ be a two-dimensional connected Riemannian manifold. We call $M$ a \emph{Riemann surface} in the following. The \emph{uniformization theorem} states that every such surface $M$ is conformally equivalent to a surface with constant curvature being either $+1$, $0$, or $-1$. (This means that, when expressed in so-called isothermal coordinates, the Riemannian metric takes the form $\mbox{d}s^2=\Omega(\textbf{x})(\mbox{d}x^2+\mbox{d}y^2)$, where $\Omega(\textbf{x})$ is such that the curvature is constant.) If $M$ is simply connected, i.e.~has "no holes", it is thus equivalent to either the sphere $\mathbb{S}^2$, the complex plane $\mathbb{C}$, or the Poincar\'{e} disk $\mathbb{D}$. If $M$ has holes, which can only happen for curvature $0$ and $-1$, then it is either equivalent to a torus with genus $g=1$ (curvature $0$), or it is a hyperbolic surface (curvature $-1$) of the form $\mathbb{D}/\Gamma$, where $\Gamma$ is a so-called Fuchsian group, introduced in Sec. \ref{SecFuchs}. If the hyperbolic surface is compact, it is fully characterized by its number of holes, which coincides with its genus $g\geq 2$. Every compact hyperbolic Riemann surface can, therefore, be thought of as a surface with at least two holes.

\begin{figure}[t]
\centering
\includegraphics[width=8cm]{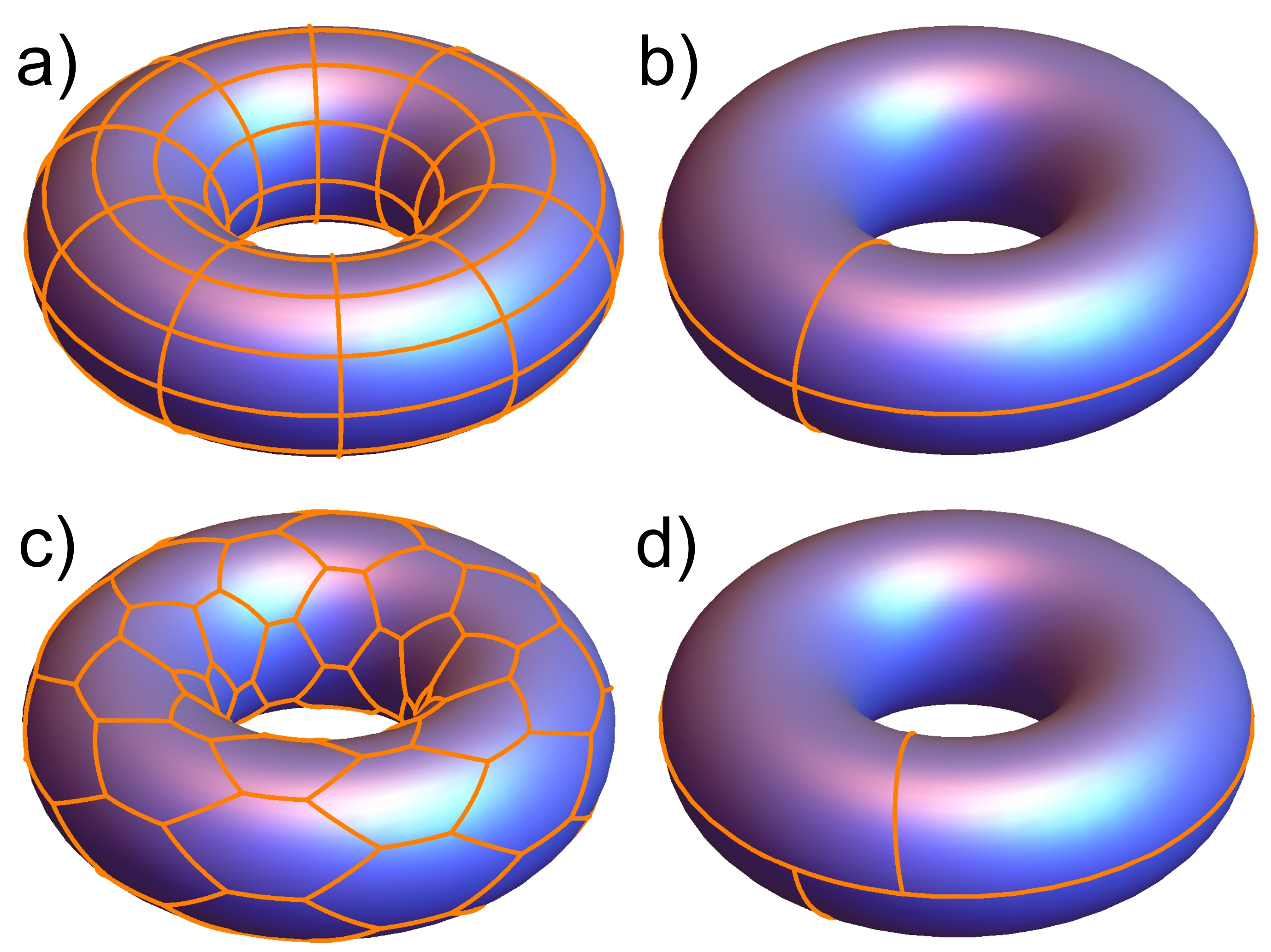}
\caption{Euclidean patterns on the torus of genus $g=1$. Figures a) and b) show $\{4,4\}$ patterns, c) and d) show $\{6,3\}$ patterns, which are tessellations of the torus by regular squares and hexagons with the appropriate coordination number. In stark contrast to hyperbolic surfaces, a torus can be tessellated with an arbitrary number of faces of an Euclidean pattern. Indeed, if $(F,E,V)$ is a solution of Eq. (\ref{pat2}) with $\chi=2(1-g)=0$, then every integer multiple thereof also has $\chi=0$ and so also can tessellate the torus. The pattern in a) is made from 100 squares and 100 vertices, the pattern in c) from 72 hexagons and 144 vertices. The patterns in b) and d) are special, because they use only one face. For the $\{4,4\}$ pattern this requires one vertex, for the $\{6,3\}$ pattern it requires two vertices, see Eq. (\ref{pat6}).}
\label{FigPat1}
\end{figure}

\begin{figure}[t]
\centering
\includegraphics[width=7cm]{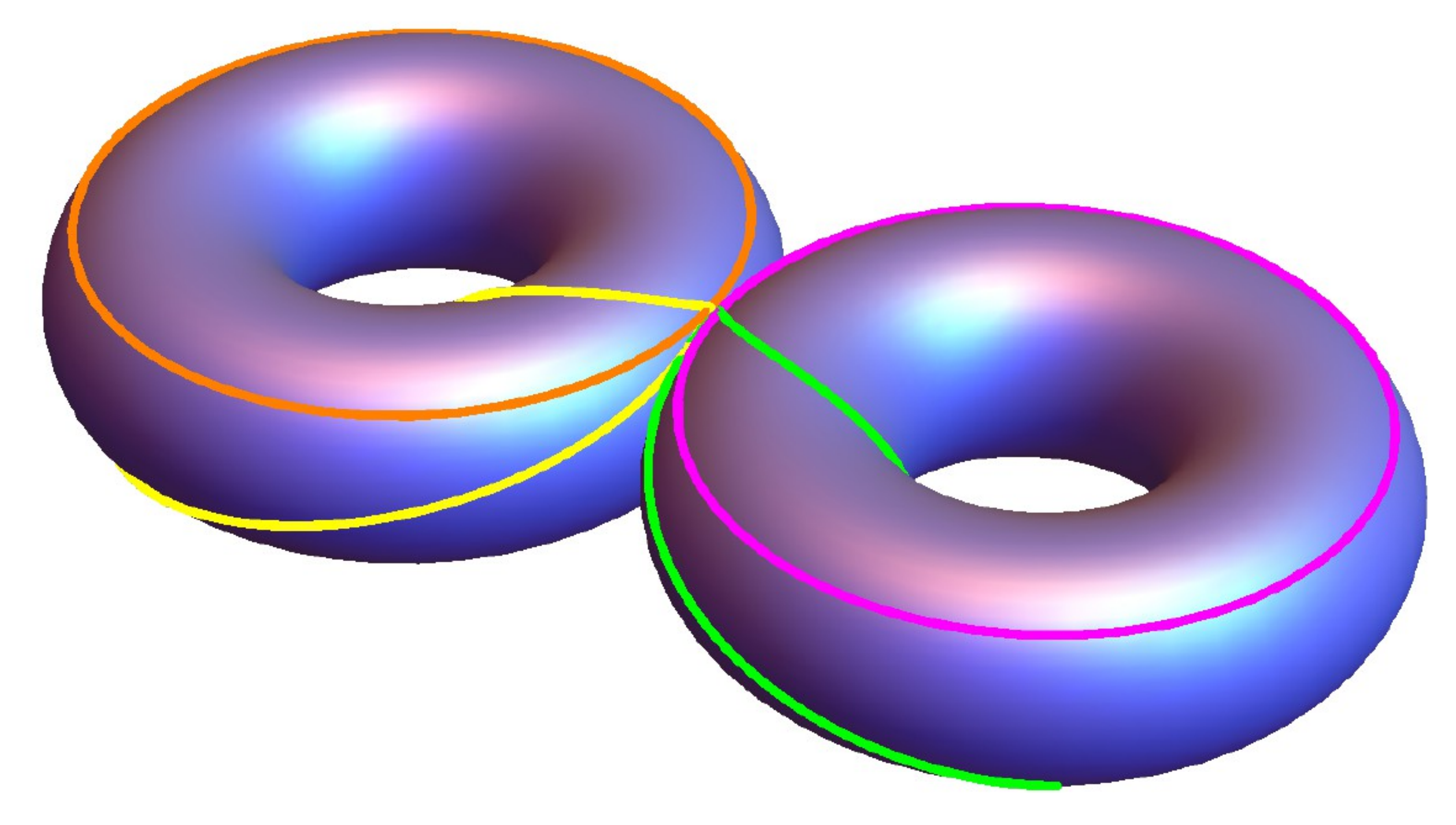}
\caption{Hyperbolic $\{8,8\}$ pattern on a surface of genus $g=2$. This pattern corresponds to the minimal solution $(F_0,E_0,V_0)=(1,4,1)$ of Eq. (\ref{pat2}) for $p=q=8$. We use four different colors to visually distinguish the four edges. We can construct this pattern by gluing together two tori from Fig. \ref{FigPat1}b) and merging the vertices from each torus into one vertex. Importantly, this pattern uses only one face. We could \emph{not} draw two faces of a $\{8,8\}$-pattern onto this surface. Indeed, since the minimal solution has $\chi=2(1-g)=-2$, the doubled solution $(2F_0,2E_0,2V_0)$ can only be embedded into a surface with $\chi=-4$ or genus $g=3$. Similar restrictions on the number of faces for a given surface of genus $g\geq 2$ apply to all hyperbolic $\{p,q\}$ patterns.}
\label{FigPat2}
\end{figure}

Following Ref. \cite{Edmonds}, we define a \emph{$\{p,q\}$ pattern on a closed Riemann surface $M$}  as a tessellation of $M$ by regular $p$-gons such that the coordination number of each vertex is $q$. Closed here means that $M$ has no boundary, and therefore such a pattern is necessarily a $q$-regular graph without boundary. Examples are shown in Figs. \ref{FigPat1} and \ref{FigPat2}. The \emph{dual pattern} is obtained by putting a vertex onto each face of the original pattern. It is easy to see that the dual pattern is then a $\{q,p\}$ pattern on $M$.

Every $\{p,q\}$ pattern on a closed surface satisfies 
\begin{align}
 \label{pat2} pF = 2E = qV,
\end{align}
where $F$, $E$, $V$ are the number of faces, edges, and vertices of the pattern, respectively. It is easy to prove this relation: Denote the adjacency matrix of the pattern by $a$, then $\sum_{i,j}a_{ij} = 2\sum_{\langle i,j\rangle} 1 = 2E$ and $\sum_{i,j}a_{ij}=q \sum_i 1 = qV$. The equality to  $pF$ follows from going to the dual graph, which has the same number of edges, but faces exchanged for vertices. The Euler characteristic $\chi$ of the pattern is given by
\begin{align}
 \label{pat3} \chi = F - E + V.
\end{align}
If $\chi$ is even, then the pattern can be embedded into an orientable surface $M$ of genus $g$ \cite{Edmonds} with 
\begin{align}
 \label{pat4} \chi =2(1-g).
\end{align}
On the other hand, if $\chi$ is odd, then the pattern can be embedded into a non-orientable surface $M$, but this case will not be of relevance to us.

\renewcommand{\arraystretch}{1.3}
\begin{table*}[t!]
\begin{center}
\begin{tabular}{|c|c|c|c||c|c|}
\hline \ $\{p,q\}$ \  & \ $F_0$ \ & \ $E_0$ \ & \ $V_0$ \ & \ $\chi_0$ \ & \ $g_0$ \ \\ 
\hline\hline $\textbf{\{6,3\}}$ & $ \textbf{1}$ & $ \textbf{3}$ & $ \textbf{2}$ & $ \textbf{0}$ & $ \textbf{1}$\\
\hline $\{7,3\}$ & $ 12$ & $ 42$ & $ 28$ & $ -2$ & $ 2$\\
\hline $\{8,3\}$ & $ 6$ & $ 24$ & $ 16$ & $ -2$ & $ 2$\\
\hline $\{9,3\}$ & $ 4$ & $ 18$ & $ 12$ & $ -2$ & $ 2$\\
\hline $\{10,3\}$ & $ 3$ & $ 15$ & $ 10$ & $ -2$ & $ 2$\\
\hline $\{11,3\}$ & $ 12$ & $ 66$ & $ 44$ & $ -10$ & $ 6$\\
\hline $\{12,3\}$ & $ 2$ & $ 12$ & $ 8$ & $ -2$ & $ 2$\\
\hline $\{13,3\}$ & \ $ 12$ \ & \ $ 78$ \ & \ $ 52$ \ & \ $ -14$ \ & $ 8$ \\
\hline $\{14,3\}$ & $ 3$ & $ 21$ & $ 14$ & $ -4$ & $ 3$\\
\hline
\end{tabular} 
\hspace{0.25cm}
\begin{tabular}{|c|c|c|c||c|c|}
\hline \ $\{p,q\}$ \  & \ $F_0$ \ & \ $E_0$ \ & \ $V_0$ \ & \ $\chi_0$ \ & \ $g_0$ \ \\ 
\hline\hline $\textbf{\{4,4\}}$ & $ \textbf{1}$ & $ \textbf{2}$ & $ \textbf{1}$ & $ \textbf{0}$ & $ \textbf{1}$\\
\hline $\{5,4\}$ & $ 8$ & $ 20$ & $ 10$ & $ -2$ & $ 2$\\
\hline $\{6,4\}$ & $ 4$ & $ 12$ & $ 6$ & $ -2$ & $ 2$\\
\hline $\{7,4\}$ & $ 8$ & $ 28$ & $ 14$ & $ -6$ & $ 4$\\
\hline $\{8,4\}$ & $ 2$ & $ 8$ & $ 4$ & $ -2$ & $ 2$\\
\hline $\{9,4\}$ & $ 8$ & $ 36$ & $ 18$ & $ -10$ & $ 6$\\
\hline $\{10,4\}$ & $ 4$ & $ 20$ & $ 10$ & $ -6$ & $ 4$\\
\hline $\{11,4\}$  & \ $ 8$ \ & \ $ 44$ \  & \ $ 22$ \  & \ $ -14$ \  & $ 8$ \\
\hline  $\textbf{\{12,4\}}$  & $ \textbf{1}$ & $ \textbf{6}$ & $ \textbf{3}$ & $ \textbf{--2}$ & $ \textbf{2}$\\
\hline
\end{tabular}
\hspace{0.25cm}
\begin{tabular}{|c|c|c|c||c|c|}
\hline \ $\{p,q\}$ \  & \ $F_0$ \ & \ $E_0$ \ & \ $V_0$ \ & \ $\chi_0$ \ & \ $g_0$ \ \\ 
\hline\hline $\{4,5\}$ & $ 10$ & $ 20$ & $ 8$ & $ -2$ & $ 2$\\
\hline $\{5,5\}$ & $ 4$ & $ 10$ & $ 4$ & $ -2$ & $ 2$\\
\hline $\{6,5\}$ & $ 5$ & $ 15$ & $ 6$ & $ -4$ & $   3$\\
\hline $\{7,5\}$ & $ 20$ & $ 70$ & $ 28$ & $ -22$ & $ 12$\\
\hline $\{8,5\}$ & $ 10$ & $ 40$ & $ 16$ & $ -14$ & $ 8$\\
\hline $\{9,5\}$ & $   20$ & $ 90$ & $ 36$ & $ -34$ & $ 18$\\
\hline  $\textbf{\{10,5\}}$  & $ \textbf{1}$ & $ \textbf{5}$ & $ \textbf{2}$ & $ \textbf{--2}$ & $ \textbf{2}$\\
\hline $\{11,5\}$ & \ $ 20$ \ & \ $ 110$ \ & \ $   44$ \ & \ $ -46$ \ &  $ 24$ \\
\hline $\{12,5\}$ & $ 10$ & $ 60$ & $ 24$ & $ -26$ & $ 14$\\
\hline
\end{tabular}
\end{center}
\caption{Selection of minimal solutions $(F_0,E_0,V_0)$ to Eq. (\ref{pat2}) with even $\chi_0=F_0-E_0+V_0$ for $q=3,4,5$. For a given $\{p,q\}$, the list gives the smallest genus $g_0$ such that a $\{p,q\}$-pattern with $F_0$ faces, $E_0$ edges and $V_0$ vertices can be drawn onto an orientable surface of genus $g_0$. Every integer multiple of $(F_0,E_0,V_0)$ also yields a $\{p,q\}$ pattern, but on a surface with higher genus. (An exception is the Euclidean case with $\chi_0=0$, where the number of squares or hexagons that can be used to cover a torus is arbitrary.) We highlight in boldface solutions that constitute a pattern with a single face, since these correspond to potential regular Bravais lattices.}
\label{TabPat}
\end{table*}
\renewcommand{\arraystretch}{1}

Given a solution $(F,E,V)$ of Eq. (\ref{pat2}), we can generate more solutions by multiplying the first solution by an arbitrary integer. Crucially, for hyperbolic $\{p,q\}$ patterns, the number of faces $F$ and the genus $g\geq 2$ of the surface are not independent. This can be understood purely algebraically or, possibly more intuitively, geometrically. 

Algebraically, multiplying a solution $(F,E,V)$ of Eq. (\ref{pat2}) by an integer $n$ yields a pattern with $(nF,nE,nV)$ on a surface with characteristic $n\chi$. For $g\geq 2$ this corresponds to a surface of higher genus than the original one. Increasing the number of faces is thus equivalent to increasing the genus for hyperbolic patterns. For Euclidean patterns, with $g=1$ and $\chi=0$, on the other hand, the number of faces is not restricted. We show some instructive examples in Figs. \ref{FigPat1} and \ref{FigPat2}.

Geometrically, it is clear that the combined area of the faces of the pattern needs to match the area of the closed Riemann surface. In the Euclidean case, the size of squares or regular hexagons is arbitrary, and a matching is possible for any number of faces. For regular hyperbolic polygons in a $\{p,q\}$ pattern, on the other hand, the area of a single polygon is fixed to $A(p,q)=(p-2)\pi-p\frac{2\pi}{q}$. The area of the hyperbolic surface is $4\pi(g-1)$ via the Gau\ss{}--Bonnet theorem, and so we arrive at the necessary condition 
\begin{align}
 \label{patX} F A(p,q)=4\pi(g-1),
\end{align}
which relates $F$ and $g$. This condition is satisfied if $(F,E,V)$ solves Eq. (\ref{pat2}).

For every $\{p,q\}$, there exists a minimal solution $(F_0,E_0,V_0)$ with smallest number of faces $F_0$ (and therefore smallest $E_0$ and $V_0$). To find the minimal solution, start with $F_0=1$ and check whether $pF_0$ is divisible by $2$ and $q$, and, if not, increase $F_0$ by one unit. We restrict admissible minimal solutions to even values of $\chi$, thus orientable surfaces, multiplying by two if the algorithm described above yields an odd $\chi$. We present a selection of minimal solutions in Table \ref{TabPat}.

Among the minimal solutions of patterns, the ones with $F_0=1$ stand out. If a $\{p,q\}$ pattern can be embedded into a closed surface with only one face, then this implies that we can consistently define periodic boundary conditions on the associated regular $p$-gon with interior angles $2\pi/q$. This connection has been explained in detail in Ref. \cite{sausset2007periodic}. Every solution with $F_0=1$ satisfies 
\begin{align}
 \label{pat5} (F_0,E_0,V_0)=(1,p/2,p/q),
\end{align}
hence $p$ must be even and must satisfy $p\geq q$. It is easy to see that four infinite families of such patterns are
\begin{align}
 \nonumber &\{4g,4g\}:\ (F_0,E_0,V_0) = (1,2g,1),\\
 \nonumber &\{2(2g+1),2g+1\}:\ (F_0,E_0,V_0) = (1,2g+1,2),\\
 \nonumber &\{4(2g-1),4\}:\ (F_0,E_0,V_0) =(1,2(2g-1),2g-1),\\
 \label{pat6} &\{6(2g-1),3\}:\ (F_0,E_0,V_0) =(1,3(2g-1),2(2g-1)),
\end{align}
where $g\geq 1$ is the genus of the embedding surface. These families obviously generalize the square and hexagonal lattices, $\{4,4\}$ and $\{6,3\}$, to higher genus. One can show that every solution with $F_0=1$ and even $\chi$ is contained in one of the two families: (i) Either $q$ is a multiple of 4, then the associated pattern is of type
\begin{align}
 \label{pat8} \{4m(2n+1),4m\},
\end{align}
or (ii) $q$ is odd, so that the pattern is of type
\begin{align}
 \label{pat9} \{2(2m+1)(2n+1),2m+1\}.
\end{align}
In both cases, $m\geq1$ and $n\geq 0$ are integers, and the genus is $g=(2n+1)m-n$. For $(m,n)=(g,0)$ and $(m,n)=(1,g-1)$, we recover the four patterns from Eq. (\ref{pat6}). For a given $g$, more than the four patterns in Eq. (\ref{pat6}) may exist. We summarize the solutions for $g=2,3$ in Table \ref{TabPatOneface}.

\renewcommand{\arraystretch}{1.3}
\begin{table}
\begin{center}
\begin{tabular}{|c||c|c|c||c|c|}
\hline \ $\{p,q\}$ \  & \ $F_0$ \ & \ $E_0$ \ & \ $V_0$ \ & \ $\chi_0$ \ & \ $g_0$ \ \\ 
\hline\hline $\{8,8\}$ & $ 1$ & $ 4$ & $ 1$ & $ -2$ & $ 2$\\
\hline $\{10,5\}$ & $ 1$ & $ 5$ & $ 2$ & $ -2$ & $ 2$\\
\hline $\{12,4\}$ & $ 1$ & $ 6$ & $ 3$ & $ -2$ & $ 2$\\
\hline $\{18,3\}$ & $ 1$ & $ 9$ & $ 6$ & $ -2$ & $ 2$\\
\hline\hline $\{12,12\}$ & $ 1$ & $ 6$ & $ 1$ & $ -4$ & $ 3$\\
\hline $\{14,7\}$ & $ 1$ & $ 7$ & $ 2$ & $ -4$ & $ 3$\\
\hline $\{20,4\}$ & $ 1$ & $ 10$ & $ 5$ & $ -4$ & $ 3$\\
\hline $\{30,3\}$ & $ 1$ & $ 15$ & $ 10$ & $ -4$ & $ 3$\\
\hline
\end{tabular} 
\end{center}
\label{TabPatOneface}
\caption{All possible $\{p,q\}$ patterns that can be drawn on orientable compact surfaces of genus $g=2,3$ with a single face (i.e.~such patterns are solutions to Eq. (\ref{pat2}) with $F_0=1$). The number of these solutions varies with $g$. For instance, there are four admissible patterns for $g=4,6$, but six solutions for $g=5$.}
\end{table}
\renewcommand{\arraystretch}{1}

\begin{figure}[t]
\centering
\includegraphics[width=8cm]{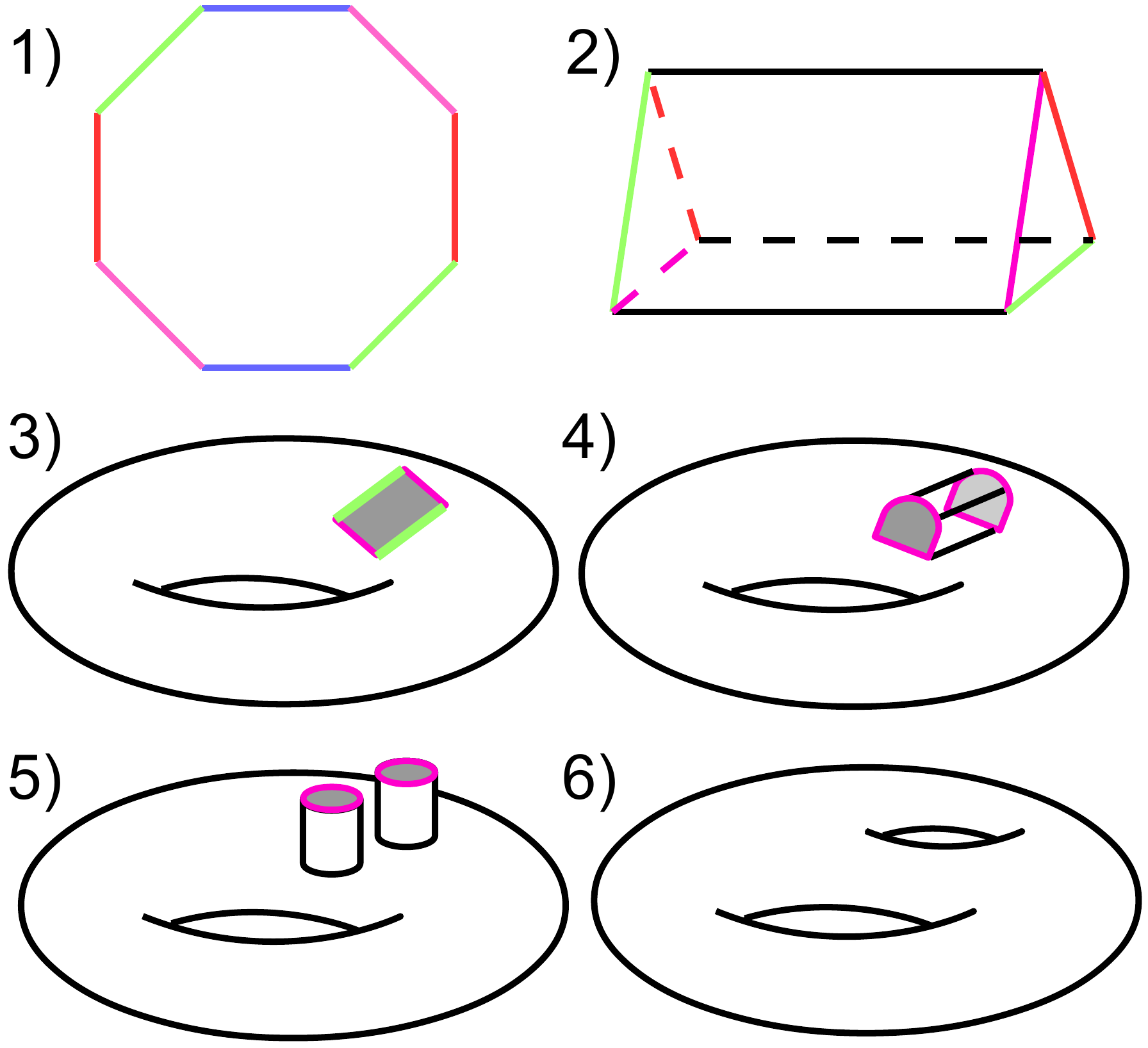}
\caption{Identifying opposite sides of an octagon yields a closed genus-2 Riemann surface. We show one possible set of steps to arrive at this well-known result. We only consider the topology and, therefore, are free to deform the surface in any way. \emph{1)} Opposite edges of the octagon which are to be identified are represented by the same color. \emph{2)} Gluing together the blue edges, we obtain a prism with triangular base. \emph{3)} Identifying the red edges, we obtain a torus with a rectangular window. The sides of the window are the yet unidentified green and pink edges. \emph{4)} Gluing together the green edges, we arrive at a torus with a tunnel on its surface. The entrance and exit of the tunnel are the pink edges. \emph{5)} Topologically, this is equivalent to a torus with two chimneys. \emph{6)} Eventually, identifying the pink rims of both chimneys, we obtain a surface with two handles.}
\label{FigMakingGenus2}
\end{figure}

The fact that a $\{4g,4g\}$ pattern can be drawn onto a surface of genus $g\geq 1$ using one face and one vertex gives an elegant way to construct higher-genus surfaces by taking a single hyperbolic $4g$-gon in the Poincar\'{e} disk (with interior angles $2\pi/(4g)$) and identifying opposite edges to obtain a closed manifold. For $g=1$, identifying opposite sides of a square yields a torus, while for $g=2$, identifying opposite sides of an octagon yields a genus-2 surface, and so on. The octagon case is visualized in Fig. \ref{FigMakingGenus2}. If we wish to equip a hyperbolic $p$-gon (with interior angles $2\pi/q$) with periodic boundary conditions by identifying certain edges, then this is possible if and only if the corresponding $\{p,q\}$ pattern can be drawn onto a closed surface using only a single face. This shows how the solutions with $F_0=1$ are crucial for identifying a periodic pattern in general hyperbolic lattices. We illuminate this setup with examples in the next sections. Finally, we note that we need not necessarily identify opposite edges of the $p$-gon. Other side-pairings are possible, but not relevant for our considerations \cite{BookMagnus,BookCoxeter}.

\section{Hyperbolic Crystallography}\label{SecCrys}

In this section, we generalize the basic notions of Euclidean crystallography to the hyperbolic case. For this purpose, we first discuss unit cells and Bravais lattices in symmorphic space groups. We formulate the translation groups in the square and hexagonal lattice in a fashion that generalizes to higher genus. We then present the discrete symmetry and translation groups for hyperbolic lattices with regular Bravais lattices of the types $\{4g,4g\}$ and $\{2(2g+1),2g+1\}$. We close this section with a remark on the order of the point group in hyperbolic lattices.

\subsection{Unit cell and Bravais lattice}\label{SecUnit}

We first recall the crystallographic notions of unit cell and Bravais lattice. For a detailed introduction we refer to Ref. \cite{BookDresselhaus}. Given a discrete set of points $\Lambda=\{z_i\}$ that constitutes the lattice, there exists a maximal group $\mathcal{G}$ acting on the coordinates that leaves the lattice invariant, called the \emph{space group}. We assume in the following that this space group is symmorphic \cite{BookDresselhaus}. The lattice can then be split into unit cells and a Bravais lattice in the following manner. Each site $z_i\in \Lambda$ can be uniquely written as 
\begin{align}
 \label{bra1} z_i = \gamma z^{(a)},
\end{align}
where $z^{(a)}$ is an element from a reference unit cell $D=\{z^{(1)},\dots,z^{(N)}\}\subset \Lambda$, which consists of a finite number of sites, and where $\gamma$ is an element from a discrete translation group $\Gamma\subset \mathcal{G}$, which is the symmetry group of the \emph{Bravais lattice}. Roughly speaking, a translation is a symmetry transformation without fixed point. The split in Eq. (\ref{bra1}) allows us to uniquely write the index of $z_i$ as $i=(\gamma,a)$, where $\gamma$ and $a$ specify the location of $z_i$ in the Bravais lattice and unit cell, respectively. For such a split to exist, $\Gamma$ needs to be a normal subgroup of $\mathcal{G}$, which follows from the assumption that $\mathcal{G}$ is symmorphic. The corresponding quotient group $G=\mathcal{G}/\Gamma$ is the \emph{point group} of the lattice.

\begin{figure}[t]
\centering
\includegraphics[width=8.5cm]{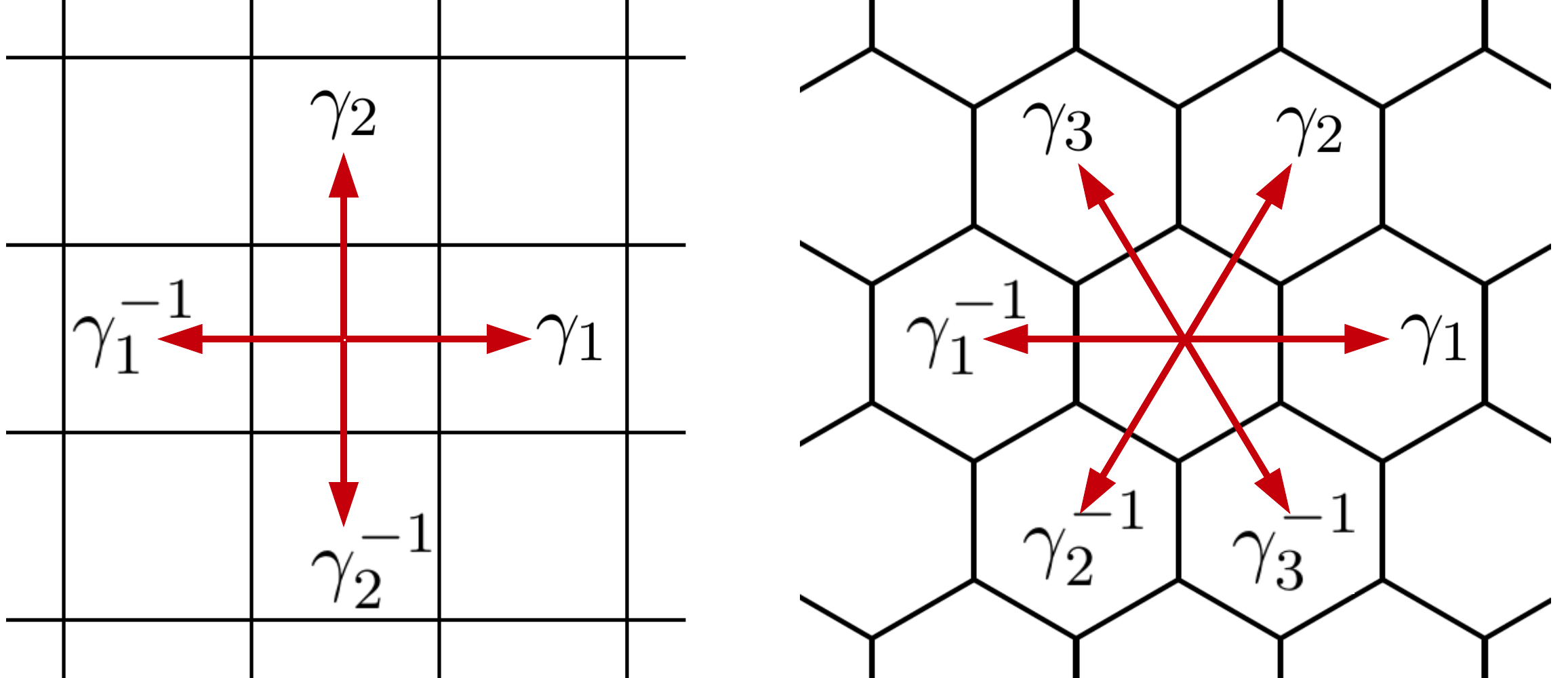}
\caption{Euclidean Bravais lattices have two independent generators of translations. This implies that momentum space in Euclidean Bloch wave theory is two-dimensional. \emph{Left.} In the $\{4,4\}$ lattice, translations are generated by the two primitive translations $\gamma_1$ and $\gamma_2$ from Eq. (\ref{bra5}). Both operations commute, which can be written as $\gamma_1\gamma_2^{-1}\gamma_1^{-1}\gamma_2=1$. The latter relation represents the fact that going (right) around a vertex four times brings one back to where one started. \emph{Right.} In the $\{6,3\}$ lattice, three generators $\gamma_1, \gamma_2, \gamma_3$ can be defined naively, see Eq. (\ref{bra9}). However, only two of them are independent, since $\gamma_1\gamma_2^{-1}\gamma_3=1$. The remaining two satisfy $\gamma_1\gamma_2^{-1}\gamma_1^{-1}\gamma_2=1$ as in the square lattice case.}
\label{FigEucGen}
\end{figure}

As an example of how to construct a lattice with a given unit cell and Bravais lattice, consider the Euclidean $\{4,4\}$ lattice in Fig. \ref{FigEucGen}. We decorate a single "fundamental" square with a finite number of sites, $D=\{z^{(1)},\dots,z^{(N)}\}$, each site placed within the same square. Since the fundamental square can cover a torus, we could endow it with periodic boundary conditions by identifying opposite sides, and the unit cell would now be embedded on a torus. On the other hand, instead of covering a torus with a single face, we may also use the fundamental square to tessellate the Euclidean plane. In this alternative point of view, when leaving one square by applying one of the primitive translation vectors of the square lattice, say $\gamma_1$, we do not enter the same square, but rather enter the \emph{neighboring} square. We iterate this translation procedure, using all sides of the fundamental domain. In this way, we tessellate the Euclidean plane with repetitions of $D$ and the resulting periodic set of sites $\{z_i\}$ is a Euclidean lattice with unit cell $D$ and Bravais lattice $\{4,4\}$.

\begin{figure}[t!]
\centering
\includegraphics[width=8.5cm]{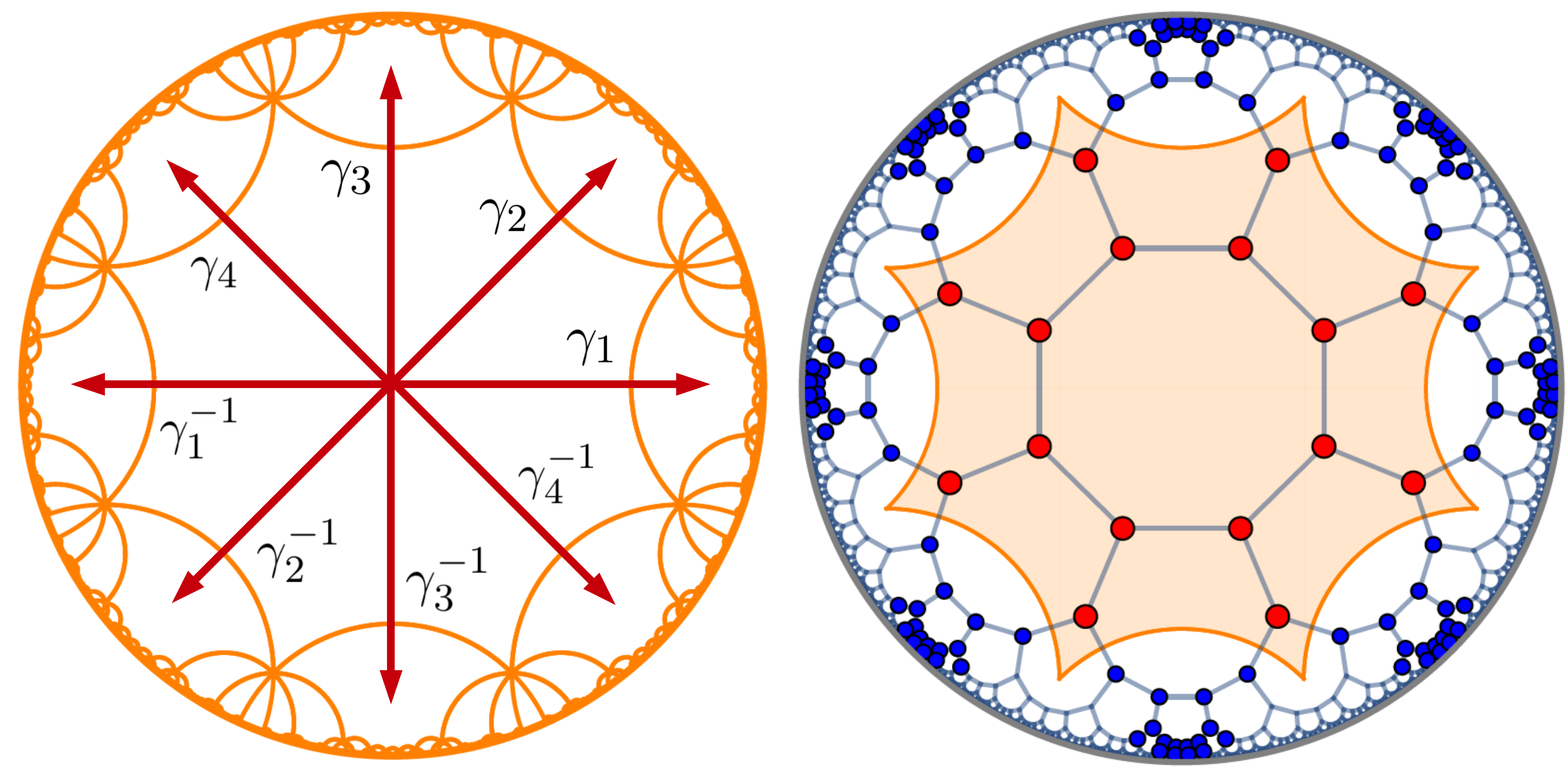}
\caption{\emph{Left.} The $\{8,8\}$ lattice, shown in orange, is a hyperbolic Bravais lattice with $g_0=2$. The eight edges of its central octagon define four generators of translations, which we call $\gamma_1, \gamma_2, \gamma_3, \gamma_4$. Each face of the $\{8,8\}$ lattice is reached from the central polygon by applying a product of the four generators and their inverses. The Euclidean analogue of this construction is shown in Fig. \ref{FigEucGen}. \emph{Right.} The unit cell of the $\{8,3\}$ lattice has 16 sites (red dots). The corresponding Bravais lattice is the $\{8,8\}$ lattice, with the fundamental octagon shown in orange. By applying on the original 16-site unit cell each of the generators $\gamma_1, \gamma_2, \gamma_3, \gamma_4$ of the $\{8,8\}$ lattice and their inverses once, we generate $8\times 16 = 128$ new sites (blue dots). Iterating this procedure we eventually generate the whole $\{8,3\}$ lattice, see Fig. \ref{FigGrow83}.}
\label{FigPatching}
\end{figure}

Clearly, if we started with a fundamental hexagon of the $\{6,3\}$ lattice in the previous example, we would have obtained a lattice in the Euclidean plane with Bravais lattice $\{6,3\}$, see Fig. \ref{FigEucGen}. It is then very natural to ask whether starting from a fundamental $p_{\rm B}$-gon we obtain a lattice whose Bravais lattice is $\{p_{\rm B},q_{\rm B}\}$ for some $q_{\rm B}$. This expectation turns out to be true, but not every pair of integers $(p_{\rm B},q_{\rm B})$ qualifies for a potential Bravais lattice. In fact, the construction described in the previous paragraph relies on assigning consistent periodic boundary conditions to the fundamental polygon. This is possible if and only if $(p_{\rm B},q_{\rm B})$ allows for a solution of Eq. (\ref{pat2}) with $F_0=1$, i.e.~the fundamental domain can cover a closed surface with a single face. Obviously, this condition is satisfied for the Euclidean examples $\{4,4\}$ and $\{6,3\}$. In the hyperbolic case, we see that only certain $\{p_{\rm B},q_{\rm B}\}$ lattices, such as the infinite families $\{4g,4g\}$ or $\{2(2g+1),2g+1\}$, constitute valid Bravais lattices. Remarkably, there are \emph{infinitely many Bravais lattices in the hyperbolic plane}. We demonstrate the construction of the $\{8,3\}$ lattice by decorating the fundamental octagon of the $\{8,8\}$ Bravais lattice in App. \ref{SecGen}. The idea is outlined in Fig. \ref{FigPatching}.

A Bravais lattice will be called \emph{regular} if its fundamental domain is a regular polygon, and so is a $\{p_{\rm B},q_{\rm B}\}$ lattice for some $p_{\rm B}$ and $q_{\rm B}$. This is a severe constraint. For instance, the internal angles of the fundamental polygon may not all be equal, while still yielding a valid Bravais lattice. We will see, however, that many examples that are important for experiments and applications fall into the class of regular Bravais lattices.

\subsection{Euclidean case}\label{SecEucl}

In the next few paragraphs, we return to the Euclidean example from the previous section and explicitly construct the translations that facilitate the tessellation of the Euclidean plane. It is well-known that Euclidean lattices in two dimensions are constructed from five Bravais lattices and 17 possible space groups, the latter called wallpaper groups in this context \cite{BookDresselhaus}. The five Bravais lattices can be characterized by their two primitve translation vectors, $\textbf{a}_1$ and $\textbf{a}_2$. Let $\theta$ denote the angle between $\textbf{a}_1$ and $\textbf{a}_2$. Among the five Bravais lattices, only the square lattice and hexagonal lattice are regular, i.e. $|\textbf{a}_1|=|\textbf{a}_2|$, with  $\theta=90^\circ$ and $\theta=120^\circ$, respectively. The remaining three (oblique, rectangular, centered rectangular) have $|\textbf{a}_1|\neq|\textbf{a}_2|$ and so are not regular. As elucidated before, the appearance of the $\{4,4\}$ and $\{6,3\}$ lattices as regular Bravais lattices in the Euclidean plane is deeply rooted in the fact that these lattices can cover a genus-one torus with a single face.

Let us explicitly construct the translation group $\Gamma$ for the $\{4,4\}$ Bravais lattice. For a $2\times 2$ matrix $M$, we define its action on $z\in \mathbb{C}$, denoted $Mz$, through the generalization of Eq. (\ref{hyp4}) by
\begin{align}
 \label{bra2} \begin{pmatrix} a & b \\ c & d \end{pmatrix} z := \frac{az+b}{cz+d},
\end{align}
assuming, of course, that $M$ is such that the denominator is nonzero. The square lattice shall be aligned as in Fig. \ref{FigEucGen}. We choose the side length or lattice constant to be $a_0=1$. Every translation of the fundamental square is generated by the maps
\begin{align}
 \label{bra3} z&\mapsto \gamma_1 z = z+1,\\
 \label{bra4} z&\mapsto \gamma_2 z = z+\rmi,
\end{align}
corresponding to the two primitive translation vectors $a_1=1$ and $a_2=\rmi$ in complex notation, and matrices
\begin{align}
 \label{bra5} \gamma_1 = \begin{pmatrix} 1 & 1 \\ 0 & 1 \end{pmatrix},\ \gamma_2 = \begin{pmatrix} 1 & \rmi  \\0  & 1 \end{pmatrix}.
\end{align}
Clearly, the two operations commute, $\gamma_1\gamma_2=\gamma_2\gamma_1$, which can be written as
\begin{align}
 \label{bra6} \gamma_1\gamma_2^{-1}\gamma_1^{-1}\gamma_2 = \mathbb{1}.
\end{align} 
We call $\gamma_1$ and $\gamma_2$ the generators of the translation group. Every translation connecting one point of the Bravais lattice to another can be written as a suitable product of $\gamma_1,\gamma_2$, and their inverses. Consequently, Eqs. (\ref{bra5}) specify a particular representation of the translation group $\Gamma$ of the $\{4,4\}$ lattice. The abstract presentation of the same group reads
\begin{align}
 \label{bra7} \Gamma_{g=1} = \langle \gamma_1, \gamma_2\ |\ \gamma_1\gamma_2^{-1}\gamma_1^{-1}\gamma_2 = 1\rangle \simeq\mathbb{Z}^2.
\end{align}
Here we use the standard notation $\langle A,B,\dots| X=Y=\dots=1\rangle$ for a group generated by some $A,B,\dots$ and their inverses that satisfy the constraints $X=Y=\dots=1$. An element of $\Gamma$, which is a certain ordered product of the generator and their inverses, is called a \emph{word}. The group $\Gamma_{g=1}$, of course, is isomorphic to $\mathbb{Z}^2$.

Next we construct the translation group $\Gamma$ for the Euclidean $\{6,3\}$ Bravais lattice. The lattice shall be aligned as in Fig. \ref{FigEucGen}. Translations through the sides of the hexagon are generated by $z\mapsto \gamma_\mu z = z+ e_\mu$, with lattice constant $a_0=1$, primitive vectors $e_1=1,\ e_2= e^{\rmi \pi/3},\ e_3 = e^{2\rmi \pi/3}$ and translation generators
\begin{align}
 \label{bra9} \gamma_1 = \begin{pmatrix} 1 & e_1 \\ 0 & 1 \end{pmatrix},\  \gamma_2 = \begin{pmatrix} 1 & e_2 \\ 0 & 1 \end{pmatrix},\  \gamma_3 = \begin{pmatrix} 1 & e_3 \\ 0 & 1 \end{pmatrix}.
\end{align}
Note that we have $e_1-e_2+e_3=0$ and so
\begin{align}
 \label{bra10} \gamma_1\gamma_2^{-1}\gamma_3 = \begin{pmatrix} 1 & (e_1-e_2+e_3) \\ 0 & 1 \end{pmatrix} = \mathbb{1}.
\end{align}
Hence the number of independent generators is two, not three, just as for the $\{4,4\}$ lattice. The translation group of the $\{6,3\}$ lattice is thus given by
\begin{align}
 \nonumber \Gamma_{\{6,3\}} &= \langle \gamma_1, \gamma_2, \gamma_3 \ |\ \gamma_1\gamma_2^{-1}\gamma_3  =\gamma_1\gamma_2^{-1}\gamma_1^{-1}\gamma_2 = 1\rangle\\
 \label{bra11}  &= \langle \gamma_1, \gamma_2 \ |\ \gamma_1\gamma_2^{-1}\gamma_1^{-1}\gamma_2 = 1\rangle = \Gamma_{g=1}.
\end{align}
Since $\gamma_3$ can be expressed in terms of $\gamma_1$ and $\gamma_2$, every word in $\gamma_{1,2,3}$ is also a word in $\gamma_{1,2}$. The two Euclidean translation groups are, therefore, isomorphic and fully characterized by the genus $g=1$.

The two Euclidean lattices discussed here are the special case of $g=1$ for the genus-$g$ lattices $\{4g,4g\}$ and $\{2(2g+1),2g+1\}$ analyzed in the following. Although similarities between the Euclidean and hyperbolic cases remain, the most striking difference for $g\geq 2$ is the fact that the generators $\gamma_\mu$ no longer commute. The proper framework to discuss these non-commuting translations is the language of Fuchsian groups.

\subsection{Fuchsian groups}\label{SecFuchs}

A discrete subgroup of $\mathcal{P}=\text{PSU}(1,1)$ is called a \emph{Fuchsian group}. It is very natural to expect symmetry groups of hyperbolic lattices to be Fuchsian groups \cite{BookMagnus,BookCoxeter,BALAZS1986109,BookBujalance}. Indeed, under a symmetry transformation of the lattice, two neighboring sites $z_i$ and $z_j$, separated by a hyperbolic distance $d(z_i,z_j)=d_0$ that is determined by $p$ and $q$, should be mapped to two neighboring points separated by the same hyperbolic distance. Since $\mathcal{P}$ is precisely the group of transformations that preserve the hyperbolic distance, the symmetry group must be made from elements of $\mathcal{P}$. On the other hand, clearly only a discrete set of transformations will leave the lattice invariant.

The full space group of the hyperbolic $\{p,q\}$ lattice is given by the \emph{triangle group}
\begin{align}
 \nonumber \Delta(p,q,2) = \langle\ x,y,z\ |\ x^2&=y^2=z^2=(xy)^p\\
 \label{bra12} &=(yz)^q=(zx)^2=1\ \rangle.
\end{align}
The geometric meaning of the generators $x,y,z$ is not important for this work. Suffice to say that this group contains a reflection along a symmetry axis and thus orientation-reversing elements, hence is not a subgroup of $\mathcal{P}$, but more generally referred to as \emph{non-Euclidean crystallographic group} \cite{BookCoxeter,BookBujalance}. On the other hand, we can consider the quotient $\Delta^+(p,q,2)=\Delta(p,q,2)/\mathbb{Z}_2$ of transformations modulo this reflection, which consists of orientation-preserving automorphisms and thus is a Fuchsian group. Equivalently, the full space group is given by the semi-direct product $\Delta(p,q,2)=\Delta^+(p,q,2)\ltimes \mathbb{Z}_2$. Due to this simple nature of the factor $\mathbb{Z}_2$, we will often ignore the reflection symmetry. We refer to $\Delta^+(p,q,2)\subset \mathcal{P}$ as \emph{proper triangle group}. It has the presentation
\begin{align}
\label{bra13} \Delta^+(p,q,2) = \langle\ A,B\ |\ A^p = B^q = (AB)^2 = 1_{\mathcal{P}}\ \rangle,
\end{align} 
with $A=xy$, $B=yz$. A particular representation of the generators $A$ and $B$ through $\mathcal{P}$-matrices is given by
\begin{align}
 \label{bra14} A&= \begin{pmatrix} e^{\rmi \alpha/2} & 0 \\ 0 & e^{-\rmi \alpha/2} \end{pmatrix},\\
 \label{bra15} B&= \frac{1}{1-r_0^2}\begin{pmatrix} e^{\rmi \beta/2}-r_0^2e^{-\rmi \beta/2} & r_0(1-e^{\rmi \beta})e^{\rmi\frac{(\alpha-\beta)}{2}} \\ r_0(1-e^{-\rmi \beta})e^{-\rmi\frac{(\alpha-\beta)}{2}} & e^{-\rmi \beta/2}-r_0^2e^{\rmi \beta/2} \end{pmatrix}
\end{align}
with $r_0$ from Eq. (\ref{hyp7}) and
\begin{align}
 \label{bra16} \alpha=\frac{2\pi}{p},\ \beta=\frac{2\pi}{q}.
\end{align}
Geometrically, $A$ is a rotation by $\alpha$ through the center of a face (here chosen to be the origin), whereas $B$ is a rotation by $\beta$ through a vertex (here chosen to be $z_1=r_0 e^{\rmi \alpha/2}$).

The elements of a Fuchsian group are classified as elliptic, parabolic, or hyperbolic if their trace is less than, equal to, or greater than 2. A typical elliptic element is given by the rotation matrix
\begin{align}
 \label{bra17} R(\phi) & = \begin{pmatrix} e^{\rmi \phi/2} & 0 \\ 0 & e^{-\rmi \phi/2} \end{pmatrix}
\end{align}
with $\phi\neq 0$. Indeed, under $R(\phi)$ we have $z\mapsto e^{\rmi \phi}z$ and $\phi$ is the angle of rotation. Elliptic elements have one fixed point, which here is the center of the rotation. A typical hyperbolic element is given by the matrix
\begin{align}
 \label{bra18} T(\tau) = \begin{pmatrix} \cosh(\tau/(2\kappa)) & \sinh(\tau/(2\kappa)) \\ \sinh(\tau/(2\kappa)) & \cosh(\tau/(2\kappa)) \end{pmatrix}
\end{align} 
with $\tau > 0$. The significance of the parameter $\tau$ can be understood from applying $T(\tau)$ to the origin $z=0$. We have
\begin{align}
 \label{bra19}  0 \mapsto T(\tau) 0 = \tanh(\tau/(2\kappa)).
\end{align}
Now note that Eq. (\ref{hyp2}) implies $d(z,0)=(2\kappa)\text{artanh}(|z|)$. Consequently, under $T(\tau)$, the origin is mapped to the coordinate on the real axis that is at hyperbolic distance $\tau$ from the origin. This finding, together with the form of $T(\tau)$ that closely resembles a Lorentz transformation, motivates us to call $T(\tau)$ a \emph{boost transformation}, and $\tau$ the boost parameter or rapidity. Importantly, like every hyperbolic element of a Fuchsian group, boosts do not have fixed points. In this sense, they generalize Euclidean translations to the hyperbolic case. The generators $A$ and $B$ can be expressed in terms of rotations and boosts via
\begin{align}
 A&= R(\alpha),\\
 B&=  R(\alpha/2)T(\tau_0) R(\beta) T(-\tau_0)R(-\alpha/2),
\end{align} 
with $\tau_0=(2\kappa)\artanh(r_0)$.

We are now in a position to characterize the translation groups associated to Bravais lattices in hyperbolic space. We define a \emph{Fuchsian translation group} $\Gamma$ as a torsion free Fuchsian group. Torsion free means that no element $\gamma$ from $\Gamma$ satisfies $\gamma^n=1$ for some suitable integer $n$. In our case, this is ensured by $\Gamma$ being strictly hyperbolic, which means that all elements are hyperbolic. Obviously, $\Delta^+(p,q,2)$ is not a Fuchsian translation group, because the generators $A$ and $B$ satisfy $A^p=1$ and $B^q=1$.

Let us pause here for a word on notation. The intrinsic properties of a $\{p,q\}$ lattice are fully specified by the integers $p$ and $q$. This includes, for instance, the value of $r_0$ in Eq. (\ref{hyp7}), the hyperbolic distance $d_0=d(z_i,z_j)$ between any two neighboring points $z_i$ and $z_j$, or the parameters of the regular geodesic $p$-gon $\{\mathcal{C}_\mu,\ \mu=1,\dots,p\}$ with internal angles $2\pi/q$ in Eq. (\ref{hyp12}). In what follows, we will discuss $\{p,q\}$ lattices and their associated $\{p_{\rm B},q_{\rm B}\}$ Bravais lattices. To distinguish these two, we denote parameters of the Bravais lattice by a subscript ${\rm B}$, which indicates that we need to replace $\{p,q\}\to\{p_{\rm B},q_{\rm B}\}$ in the corresponding formula.

It is rather easy to construct a representation of the Fuchsian translation group $\Gamma$ for the regular $\{p_{\rm B},q_{\rm B}\}$ Bravais lattice. We restrict ourselves to the $\{4g,4g\}$ and $\{2(2g+1),2g+1\}$ Bravais lattices in the following. From Eq. (\ref{pat2}) it follows that solutions with $F_0=1$ necessarily have even $p_{\rm B}$. The fundamental domain is a regular $p_{\rm B}$-gon, with internal angles $\beta_{\rm B}$, and with edges parametrized by $\mathcal{C}_{\mu,{\rm B}}$ from Eq. (\ref{hyp12}). (An example of the fundamental polygon of the $\{8,8\}$ Bravais lattice is shown in Fig. \ref{FigTrig}.) We center the fundamental polygon at the origin and align it such that $\mathcal{C}_{1,\rm B}$ intersects the positive real axis orthogonally. Opposite sides of the polygon are identified. The first generator of $\Gamma$, $\gamma_1$, is a boost that translates one fundamental polygon through side $\mathcal{C}_{1,\rm B}$ to the neighboring polygon on the right. Consequently, the transformation
\begin{align}
 \label{bra20} \gamma_1=T(\tau_1)
\end{align}
shifts the center of the original polygon to the center of the neighboring polygon, which again lies on the real axis. Hence the boost parameter is given by $\tau_1 = 2A_{\rm B}$ with $A_{\rm B}$ from Eq. (\ref{hyp8}). This yields the explicit form
\begin{align}
 \label{bra21b} \gamma_1 &= \frac{1}{\sqrt{1-\sigma^2}}\begin{pmatrix} 1&\sigma \\ \sigma & 1 \end{pmatrix}
\end{align}
with $\sigma =\sqrt{(\cos \alpha_{\rm B}+\cos\beta_{\rm B})/(1+\cos \beta_{\rm B})}$ and $\alpha_{\rm B} = 2\pi/p_{\rm B}$, $\beta_{\rm B} = 2\pi/q_{\rm B}$. The full set of generators $\gamma_\mu$ results from conjugating this boost with a rotation by $\alpha_{\rm B}$ and reads
\begin{align}
 \label{bra22} \gamma_{\mu} = R((\mu-1)\alpha_{\rm B})\gamma_1 R(-(\mu-1)\alpha_{\rm B})
\end{align}
with $\mu=1,\dots,p_{\rm B}/2$. Since $p_{\rm B}$ is even, the generators are well-defined.

The Fuchsian translation group of the regular Bravais lattice is given by
\begin{align}
 \label{bra23} \Gamma_{\{p_{\rm B},q_{\rm B}\}} = \langle \gamma_1,\dots,\gamma_{p_{\rm B}/2}\ |\ X_{\{p_{\rm B},q_{\rm B}\}}=1_{\mathcal{P}}\rangle,
\end{align}
with $X_{\{p_{\rm B},q_{\rm B}\}}$ a constraint. For practical purposes, this constraint is often unimportant, since the representation of the generators in terms of the matrices $\gamma_\mu$ in Eq. (\ref{bra22}) automatically satisfies the constraint. We derive this constraint in Appendix \ref{AppFuchs}. The important outcomes of this analysis are (i) that only $2g$ of the generators are independent and (ii) that $X_{\{p_{\rm B},q_{\rm B}\}}$ only depends on $g$ and is given by 
\begin{align}
 \label{bra23b} X_g = \gamma_1\gamma_2^{-1} \cdots \gamma_{2g-1}\gamma_{2g}^{-1} \gamma_1^{-1}\gamma_2\cdots \gamma_{2g-1}^{-1}\gamma_{2g}.
\end{align}
Intuitively, this condition means that going (right) around a vertex $4g$ times brings one back to where one started. Hence, the Fuchsian translation group 
\begin{align}
 \label{bra23c} \Gamma_g = \langle\ \gamma_1,\dots,\gamma_{2g}\ |\ X_{g}=1_{\mathcal{P}}\ \rangle
\end{align}
is fully determined by the genus $g=g_0$ of the Bravais lattice.

\subsection{Point groups}\label{SecPoint}

Having identified the Fuchsian translation group $\Gamma$, the point group is given by $G=\Delta(p,q,2)/\Gamma$, with $\Delta(p,q,2)$ from Eq. (\ref{bra12}) being the full space group of the hyperbolic lattice. Since $\Delta(p,q,2)$ has a trivial $\mathbb{Z}_2$ factor, the same is true for $G$, and we could define the point group of orientation-preserving transformations $G^+=G/\mathbb{Z}_2$. Since this is not common in crystallography, however, we work with $G$ in the following. We denote the order (number of elements) of $G$ by $|G|$. The number $|G|=|\Delta(p,q,2)/\Gamma|$ is also called the index of $\Gamma$ in $\Delta(p,q,2)$.

In mathematical terms, $\Gamma_g$ is the surface group of a closed Riemann surface $M$ with genus $g$, i.e.~it is isomorphic to the first homotopy group of the surface, $\Gamma_g \simeq \pi_1(M)$. The order of the point group then follows from the following proposition (8.3) of Ref. \cite{Edmonds}. \emph{Let $P$ be a $\{p,q\}$-pattern on a closed surface $M$, and let $\Gamma \subset \Delta(p,q,2)$ be an associated subgroup such that $\Gamma \simeq \pi_1(M)$. Then the index of $\Gamma$ in $\Delta(p,q,2)$ is $|G|=2pF$, where $F$ is the number of faces of $P$.} This allows us to determine the size of the point group, which, since the number of finite groups of certain size is limited, often determines the point group $G$ and closed surface $M$. The factor of $2$ corresponds to the $\mathbb{Z}_2$-factor in $G=G^+\ltimes \mathbb{Z}_2$. Examples of point groups that arise for hyperbolic lattices are given in the next section.

\section{Regular Hyperbolic Bravais Lattices}\label{SecReg}

In this section, we discuss hyperbolic $\{p,q\}$ lattices with regular $\{p_{\rm B},q_{\rm B}\}$ Bravais lattices. Such a discussion involves, for a given suitable $p$ and $q$, specifying the location of sites in the unit cell and the corresponding integers $p_{\rm B}$ and $q_{\rm B}$. The size of the unit cell is denoted by $N$. 

We limit the presentation to those cases where the Bravais lattice is either of the form $\{4g,4g\}$ or $\{2(2g+1),2g+1\}$, and we call $g$ the \emph{genus of the Bravais lattice} for short. (More accurately, however, it is the genus of the closed Riemann surface that can embed the unit cell with periodic boundary conditions.) Within this restricted set, several remarkable infinite families arise that are relevant for experiments with hyperbolic lattices and applications such as hyperbolic band theory. Furthermore, for $g=2$ and $g=3$, we show that besides the members of these infinite families, a few exceptional cases exist, a behavior that potentially extends to higher genera.

The examples collected in this section have been found through a systematic search, but via a case-by-case study. Hence, although we believe that the list within the restrictions specified is rather complete, we cannot exclude that we missed outliers. Given the novelty of having a list of experimentally relevant examples, we believe that such a potential incompleteness can be tolerated. Our search method is described in Appendix \ref{AppSearch}. For future work, it will be exciting to specify criteria which are both necessary and sufficient for a $\{p,q\}$ lattice to have a regular Bravais lattice of the above kind.

\subsection{Infinite families}

We first discuss five infinite families of $\{p,q\}$ lattices and their according regular Bravais lattices. They are constructed from systematically placing unit cell sites in the fundamental polygons of either the $\{4g,4g\}$ or the $\{2(2g+1),2g+1\}$ Bravais lattices. Here systematically means that the placing naturally extends to arbitrarily large genus $g$. Furthermore, the well-known Euclidean cases are recovered in the limit $g=1$. The five families are listed in Table \ref{TabUnits2}. We have explicitly verified the entries in this table for all $g \leq 8$, which is much beyond what is experimentally relevant. It is very plausible that their construction applies to $g>8$. Therefore, we will continue to call these families "infinite".

\renewcommand{\arraystretch}{1.3}
\begin{table}
\begin{center}
\begin{tabular}{|c|c|c|}
\hline \ $\{p,q\}$ \  & \ $\{p_{\rm B},q_{\rm B}\}$ \  & \ $N$ \ \\
\hline\hline \ $\{4g,4g\}$ \  & \ $\{4g,4g\}$  \  & \ $1$ \ \\
\hline \ $\{2g+1,2(2g+1)\}$ \  & \ $\{2(2g+1),2g+1\}$  \  & \ $1$ \ \\
\hline \ $\{2(2g+1),2g+1\}$ \  & \ $\{2(2g+1),2g+1\}$  \  & \ $2$ \ \\
\hline\ $\{4g,4\}$ \  & \ $\{4g,4g\}$ \  & $2g$ \  \\
\hline \ $\{2(2g+1),3\}$ \  & \ $\{2(2g+1),2g+1\}$  \  & \ $2(2g+1)$ \  \\
\hline
\end{tabular} 
\end{center}
\caption{We identify five infinite families of $\{p,q\}$ lattices whose Bravais lattices are regular $\{4g,4g\}$ or $\{2(2g+1),2g+1\}$ lattices. The number of corresponding sites in the unit cell is denoted by $N$. The construction of these families, namely the placement of the unit cell inside the fundamental domain of the Bravais lattice, is shown in Figs. \ref{FigUnits3} and \ref{FigUnits2}.}
\label{TabUnits2}
\end{table}
\renewcommand{\arraystretch}{1}

\begin{figure}[t!]
\centering
\includegraphics[width=8.6cm]{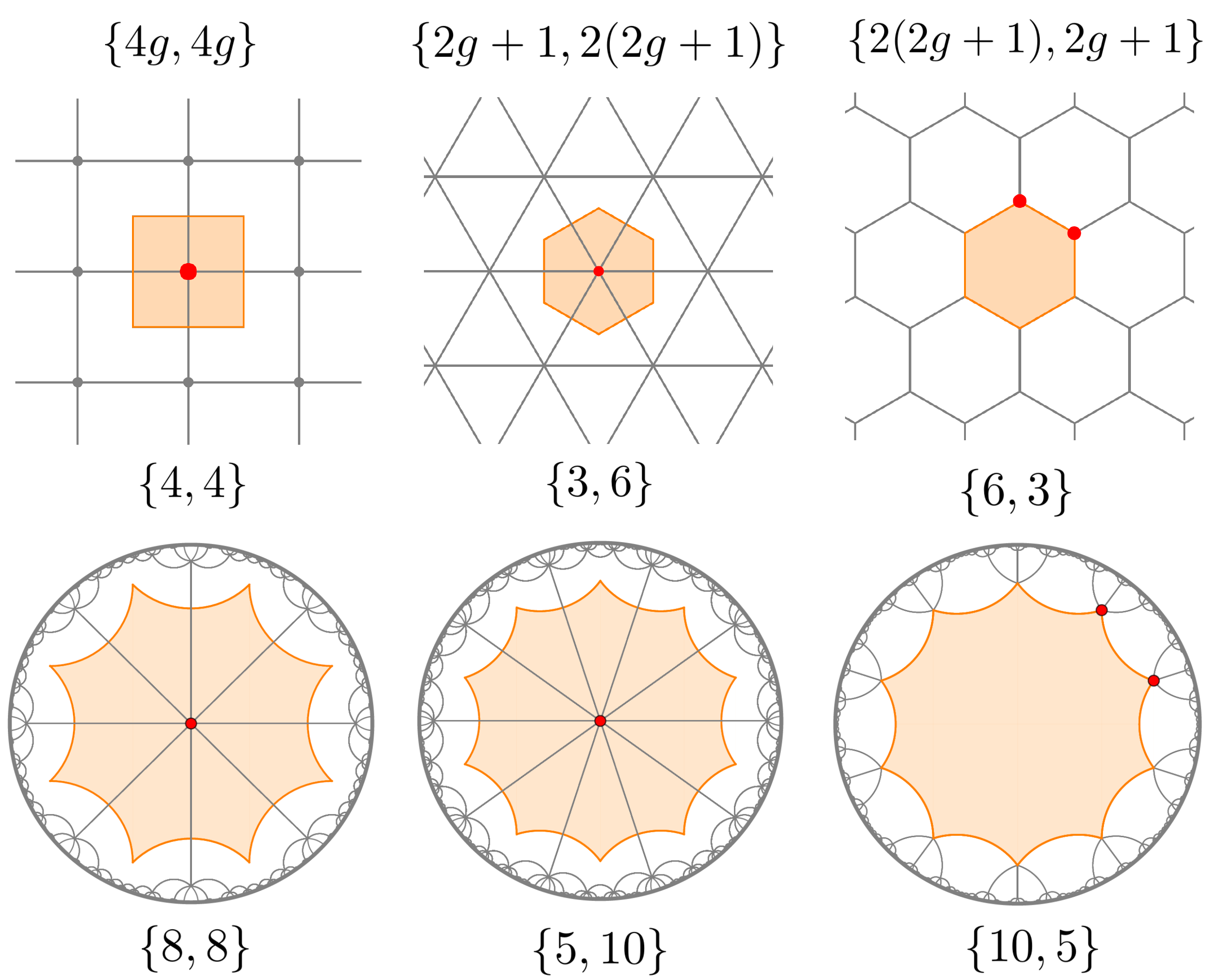}
\caption{The $\{4g,4g\}$ and $\{2(2g+1),2g+1\}$ Bravais lattices give rise to the first three infinite families in Table \ref{TabUnits2} in a simple fashion. By placing a single site into the center of the fundamental polygon, we generate the dual $\{4g,4g\}$ (left column) and $\{2g+1,2(2g+1)\}$ lattices (middle column) with a unit cell of size $N=1$. Furthermore, the $\{2(2g+1),2g+1\}$ lattice is generated from the $\{2(2g+1),2g+1\}$ Bravais lattice by placing the two unit cell sites onto vertices of the fundamental polygon (right column). All three constructions directly generalize the Euclidean case ($g=1$) to higher genus.}
\label{FigUnits3}
\end{figure}

The first family is obtained by placing a single unit cell site ($N=1$) into the center of each face of the $\{4g,4g\}$ Bravais lattice, see Fig. \ref{FigUnits3}. Applying the Fuchsian translation group to this single site, we generate the dual lattice of the Bravais lattice, which is again the $\{4g,4g\}$ lattice.  In the Euclidean case, this generates the $\{4,4\}$ lattice from the $\{4,4\}$ regular Bravais lattice. The size of the point group, due to $F_0=1$ for $\{p,q\}=\{4g,4g\}$, is $|G|=2p$. The corresponding point group is the dihedral group $D_p$, consisting of rotations by $2\pi/p$ and reflections along a symmetry axis.

The second family is obtained in a similar fashion by placing a single unit cell site ($N=1$) into the center of each face of the $\{2(2g+1),2g+1\}$ Bravais lattice. The $\{p,q\}$ lattice obtained in this manner is the $\{2g+1,2(2g+1)\}$ lattice, see Fig. \ref{FigUnits3}. For $g=1$, this construction generates the triangular $\{3,6\}$ lattice from the $\{6,3\}$ lattice. Since $\{p,q\}=\{2g+1,2(2g+1)\}$ implies $F_0=2$, the size of the point group is $|G|=4p$, and the point group is the dihedral group $D_{2p}$.

The third family is generated by placing two unit cell sites ($N=2$) on two neighboring vertices of the fundamental polygon of the $\{2(2g+1),2g+1\}$ Bravais lattice. Applying the Fuchsian translation group, we arrive at the $\{2(2g+1),2g+1\}$ lattice, see Fig. \ref{FigUnits3}. In the Euclidean example, we generate the hexagonal $\{6,3\}$ lattice from the $\{6,3\}$ lattice. The point group of the $\{p,q\}=\{2(2g+1),2g+1\}$ lattice is, again, the dihedral group $D_p$. It is well-known that the triangular lattice is the sublattice of the hexagonal lattice. How this result generalizes to the hyperbolic case is discussed in App. \ref{AppSub}.

\begin{figure}[t!]
\centering
\includegraphics[width=8.6cm]{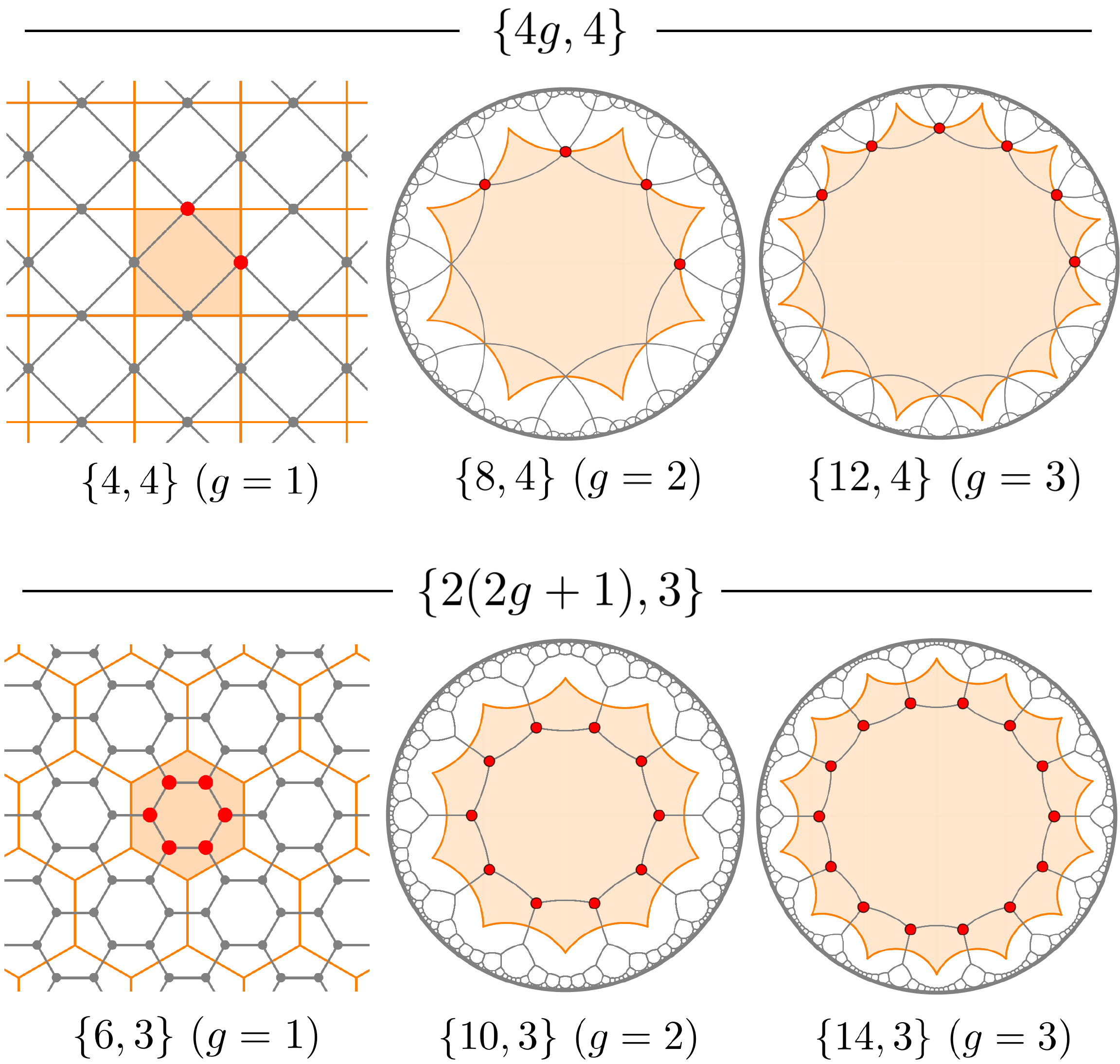}
\caption{\emph{Top row.} The infinite family of $\{4g,4\}$ lattices has a unit cell of $2g$ sites (red dots), which are placed in the edge centers of the fundamental polygon of their $\{4g,4g\}$ Bravais lattice. If $g$ is odd we can write $g=2g'-1$. In this case, the lattices are themselves $\{4(2g'-1),4\}$ Bravais lattices of genus $g'$ and a smaller unit cell can be identified. For even $g$, on the other hand, such a reduction is not possible. \emph{Bottom row.} The infinite family of $\{2(2g+1),3\}$ lattices has a unit cell of $2(2g+1)$ sites (red dots) that are facing the edges of the fundamental polygon of their $\{2(2g+1),2g+1\}$ Bravais lattice. If $2g+1$ is a multiple of 3, we can write $2g+1=3(2g'-1)$ and the lattices are $\{6(2g'-1),3\}$ Bravais lattices with a reduced unit cell and smaller fundamental domain. For other values of $g$, again, such a reduction is not possible.}
\label{FigUnits2}
\end{figure}

The fourth family generalizes the Euclidean case in a way less obvious than the previous examples. Given the fundamental $4g$-gon of the $\{4g,4g\}$ Bravais lattice, we place $N=2g$ unit cell sites on the centers of its first $2g$ edges, see Fig. \ref{FigUnits2}. This generates the $\{4g,4\}$ lattice with coordination number $4$. For $g=1$, we obtain the $\{4,4\}$ lattice from the $\{4,4\}$ Bravais lattice, mutually rotated by an angle of $\pi/4$, with a non-minimal unit cell. Indeed, in the Euclidean case, we could identify the smaller squares as the fundamental domain and so obtain a unit cell with one site. This is because the $\{4,4\}$ lattice is itself a Bravais lattice. In the hyperbolic case with $g>1$, the $\{4g,4\}$ lattice is a Bravais lattice only if $g$ is odd so that we can write $g=2g'-1$. In this case, we can identify a smaller fundamental domain for a Bravais lattice of genus $g'$, as is explained in the caption of Fig. \ref{FigUnits2}, whereas for even $g$ this is not possible. The order of the point group is $|G|=2pF_0$ with $F_0=1$ ($F_0=2$) for $g$ odd (even). In the first case, the point group is the dihedral group $D_p$.

The fifth family is obtained from the $\{2(2g+1),2g+1\}$ Bravais lattice. By placing $N=2(2g+1)$ unit cell sites, at radius $r_0$, facing the edges of the fundamental polygon of the Bravais lattice, we obtain the $\{2(2g+1),3\}$ lattice with coordination number 3, see Fig. \ref{FigUnits2}. The $\{10,3\}$ lattice discussed in Fig. \ref{Fig103and105} falls into this family. The Euclidean case corresponds to placing a smaller hexagonal lattice into a larger hexagonal Bravais lattice, and, again, in this case a smaller unit cell can be identified because the $\{6,3\}$ lattice is itself a Bravais lattice. In the hyperbolic case, as is explained in the caption of Fig. \ref{FigUnits2}, this is only possible if $2g+1$ is a multiple of 3. The order of the point group is $|G|=2pF_0$ with $F_0=1$ ($F_0=3$) for $g$ a multiple of three (else). For $F_0=1$, the point group is the dihedral group $D_p$.

\subsection{Exceptional cases for $g=2$ and $g=3$}

The list of all $\{p,q\}$ lattices for $g=2$ and $g=3$ with a regular Bravais lattice of type $\{4g,4g\}$ or $\{2(2g+1),2g+1\}$ is presented in Table \ref{TabUnits}. The five infinite families from Table \ref{TabUnits2} yield ten of the entries, but four entries are exceptional, because they do not fall into the infinite families. In the future, it will be important to study the "exceptional" cases for $g>3$ and see if they generalize to infinite families or whether they are genuinely exceptional.

\renewcommand{\arraystretch}{1.3}
\begin{table}
\begin{center}
\begin{tabular}{|c||c||c|c||c|c|}
\hline \ $\{p,q\}$ \  & \ $\{p_{\rm B},q_{\rm B}\}$ \ & \ $V_0$ \ & \ $g_0$ \  & \ $N$ \ & $g$ \\
\hline\hline \ $\{8,3\}$ \  & \ $\{8,8\}$ \ & \ $16$ \ & \ $2$ \  & \ $16$ \ & \ $2$ \ \\
\hline \ $\{8,4\}$ \  & \ $\{8,8\}$ \ & \ $4$ \ & \ $2$ \  & \ $4$ \ & \ $2$ \ \\
\hline \ $\{4,8\}$ \  & \ $\{8,8\}$ \ & \ $2$ \ & \ $2$ \  & \ $2$ \ & \ $2$ \ \\
\hline \ $\{8,8\}$ \  & \ $\{8,8\}$ \ & \ $1$ \ & \ $2$ \  & \ $1$ \ & \ $2$ \ \\
\hline\hline \ $\{10,3\}$ \  & \ $\{10,5\}$ \ & \ $10$ \ & \ $2$ \  & \ $10$ \ & \ $2$ \ \\
\hline \ $\{10,5\}$ \  & \ $\{10,5\}$ \ & \ $2$ \ & \ $2$ \  & \ $2$ \ & \ $2$ \ \\
\hline \ $\{5,10\}$ \  & \ $\{10,5\}$ \ & \ $1$ \ & \ $2$ \  & \ $1$ \ & \ $2$ \ \\
%\hline\hline \ $\{12,4\}$ \  & \ $\{12,4\}$ \ & \ \textbf{3} \ & \ $2$ \  & \ \textbf{1} \ & \ $2$ \ \\
%\hline\hline \ $\{3,9\}$ \  & \ $\{18,3\}$ \ & \ \textbf{4} \ & \ $2$ \  & \ \textbf{1} \ & \ $2$ \ \\
%\hline \ $\{18,3\}$ \  & \ $\{18,3\}$ \ & \ \textbf{6} \ & \ $2$ \  & \ \textbf{2} \ & \ $2$ \ \\
\hline\hline  \ $\{12,4\}$ \  & \ $\{12,12\}$ \ & \ $3$ \ & \ $2$ \  & \ $6$ \ & \ $3$ \ \\
\hline\ $\{4,12\}$ \  & \ $\{12,12\}$ \ & \ $1$ \ & \ $2$ \  & \ $2$ \ & \ $3$ \ \\
\hline \ $\{12,12\}$ \  & \ $\{12,12\}$ \ & \ $1$ \ & \ $3$ \  & \ $1$ \ & \ $3$ \ \\
\hline\hline \ $\{7,3\}$ \  & \ $\{14,7\}$ \ & \ $28$ \ & \ $2$ \  & \ $56$ \ & \ $3$ \ \\
\hline \ $\{14,3\}$ \  & \ $\{14,7\}$ \ & \ $14$ \ & \ $3$ \  & \ $14$ \ & \ $3$ \ \\
\hline \ $\{14,7\}$ \  & \ $\{14,7\}$ \ & \ $2$ \ & \ $3$ \  & \ $2$ \ & \ $3$ \ \\
\hline \ $\{7,14\}$ \  & \ $\{14,7\}$ \ & \ $1$ \ & \ $3$ \  & \ $1$ \ & \ $3$ \ \\
\hline
\end{tabular} 
\end{center}
\caption{List of hyperbolic $\{p,q\}$ lattices with regular $\{4g,4g\}$ or $\{2(2g+1),2g+1\}$ Bravais lattices of genus $g=2,3$. The number of unit cell sites is denoted by $N$. The values of $V_0$ and $g_0$ correspond to the minimal solution of Eq. (\ref{pat2}) for given $(p,q)$, thus $(N,g-1)$ is an integer multiple of $(V_0,g_0-1)$. The unit cells and fundamental domains of these lattices are shown in Figs. \ref{FigUnits3}, \ref{FigUnits2}, and \ref{FigUnits}.}
\label{TabUnits}
\end{table}
\renewcommand{\arraystretch}{1}

\begin{figure}[t!]
\centering
\includegraphics[width=8.5cm]{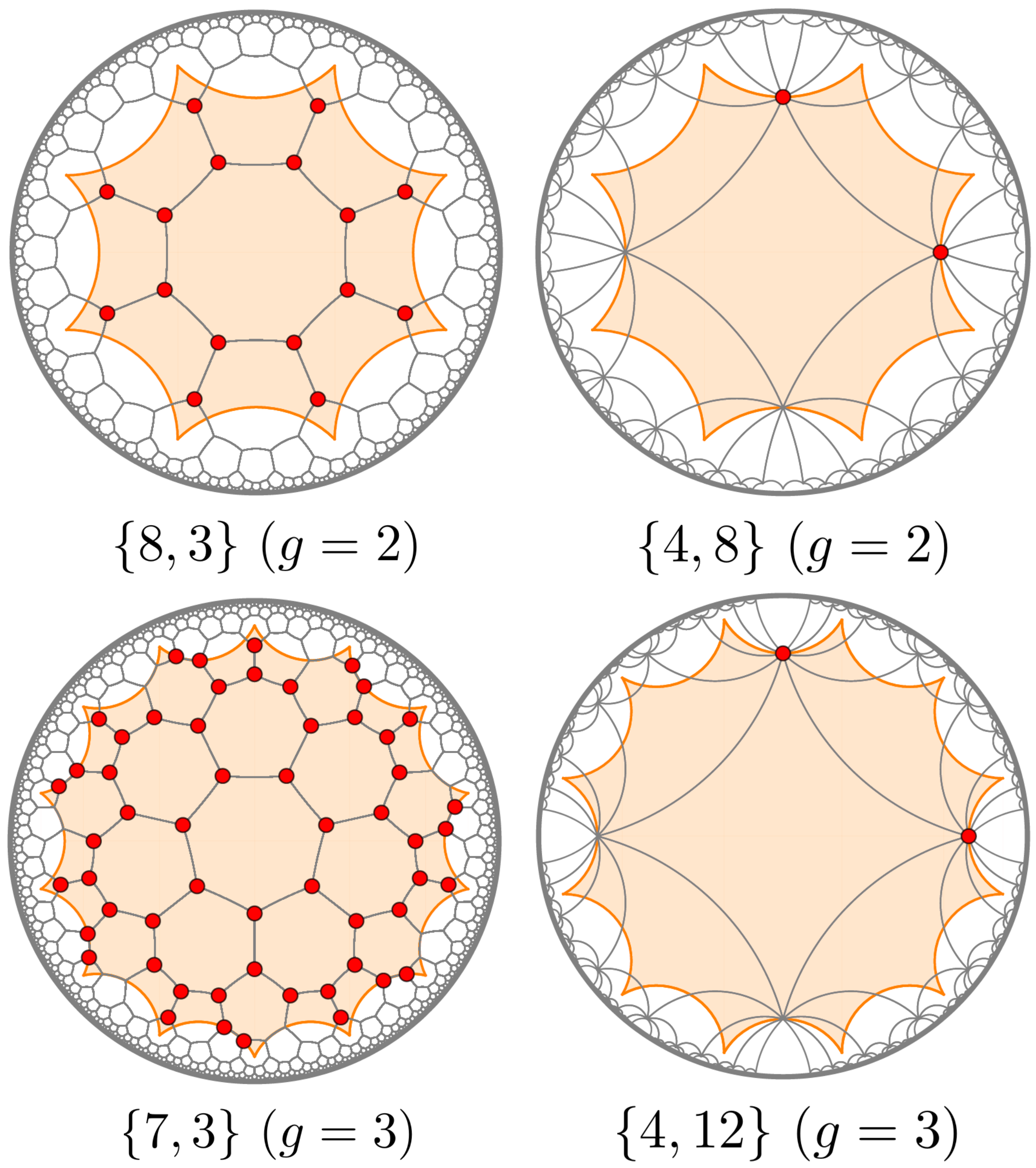}
\caption{Among the 14 hyperbolic lattices listed in Table \ref{TabUnits}, four do not fall into one of the five infinite families from Table \ref{TabUnits2}. These four exceptional cases are shown here together with their unit cell (red dots) and fundamental domain of the Bravais lattice (orange polygon). The $\{8,3\}$ lattice and its unit cell have also been discussed in Fig. \ref{FigPatching}.}
\label{FigUnits}
\end{figure}

For Bravais lattices of genus $g=2$, the $\{8,3\}$ and $\{4,8\}$ lattices are exceptional, see Fig. \ref{FigUnits}. The $\{8,3\}$ lattice, as also discussed in Fig. \ref{FigPatching}, has a 16-site unit cell inside the fundamental octagon of the $\{8,8\}$ Bravais lattice. The number of unit cell sites matches $V_0=16$ obtained from Eq. (\ref{pat2}) for $(p,q)=(8,3)$. The order of the point group follows from $F_0=6$ to be $|G|=2\cdot 48$. The ensuing pattern tessellates the so-called Bolza surface and the point group coincides with the full automorphism group of the latter, which is known explicitly. The $\{4,8\}$ lattice has a unit cell of two sites that are placed on specific edges of the fundamental octagon of the $\{8,8\}$ Bravais lattice. Again, the number of unit cell sites matches the prediction $V_0=2$ from Eq. (\ref{pat2}) for $(p,q)=(4,8)$. The size of the point group with $F_0=4$ is $|G|=2\cdot 16$.

The exceptional cases for $g=3$ Bravais lattices are the $\{7,3\}$ and $\{4,12\}$ lattices, see Fig. \ref{FigUnits}. The unit cell of the $\{7,3\}$ features 56 sites. The underlying Bravais lattice is the $\{14,7\}$ lattice. Note that the number of unit cell sites is twice the value of $V_0=28$ obtained for $(p,q)=(7,3)$ from Eq. (\ref{pat2}), and accordingly, $g=3$ is larger than the minimal solution $g_0=2$. The embedding Riemann surface is the so-called Klein quartic with full automorphism group of order $|G|=2\cdot 168$, which coincides with the point group here due to $F=24$. For the $\{4,12\}$ lattice, the two unit cell sites are located on specific edges of the fundamental dodecagon of the $\{12,12\}$ Bravais lattice. Again, the number of unit cell sites is twice the value of $V_0=1$ obtained from Eq. (\ref{pat2}) from $(p,q)=(4,12)$. The size of the point group follows from $F=6$ and is given by $|G|=2\cdot 24$.

\section{Energy spectra of tight-binding Hamiltonians}\label{SecAllApps}

In this section, we apply the crystallographic division of hyperbolic lattices into unit cells and Bravais lattices to address the problem of determining the energy spectra of tight-binding Hamiltonians. After discussing the general case, we specify to Bloch waves and the computation of their energy bands. The construction crucially relies on the possibility to \emph{uniquely} write a lattice site $z_i$, as in Eq. (\ref{bra1}), as a product of a Fuchsian translation, $\gamma \in \Gamma_g$, and a site from a reference unit cell, $z^{(a)}\in\{z^{(1)},\dots,z^{(N)}\}$, namely 
\begin{align}
\label{app1} z_i = \gamma z^{(a)}.
\end{align}
A canonical way to enumerate the infinite, but discrete, set of Fuchsian translations $\gamma$ is described in App. \ref{SecGen}.

\subsection{Tight-binding Hamiltonians on infinite lattices}\label{SecTight}

In this section, we apply hyperbolic crystallography to dramatically simplify the spectral problem for the tight-binding Hamiltonian $\hat{H}$ from Eq. (\ref{intro1}) on \emph{infinite} hyperbolic lattices. The Schr\"{o}dinger equation in coordinate representation is given by
\begin{align}
 \label{tight1} -\sum_j A_{ij} \psi(z_j) = E \psi(z_i)
\end{align}
for every site $i$, with the sum on the left extending over all lattice sites of the infinite lattice, and $A_{ij}$ the adjacency matrix. We introduce a function $A(z,z')$ such that
\begin{align}
 \label{tight2}  A_{ij}=A(z_i,z_j).
\end{align}
In the following, we consider an infinite hyperbolic $\{p,q\}$ lattice, denoted $\Lambda$, with regular $\{p_{\rm B},q_{\rm B}\}$ Bravais lattice being either the $\{4g,4g\}$ or $\{2(2g+1),2g+1\}$ lattice.

We divide $\Lambda$ into patches of unit cells, each surrounded by a fundamental polygon of the Bravais lattice, and choose a reference unit cell $D=\{z^{(1)},\dots,z^{(N)}\}\subset \Lambda$. In Eq. (\ref{tight1}), we write $z_i = \gamma z^{(a)}$ as in Eq. (\ref{app1}). Furthermore, every neighboring site of $z_i$ must be of the form $\gamma\gamma' z^{(b)}$, with some $\gamma'\in\Gamma$ and $b\in D$. Hence we arrive at
\begin{align}
 \label{tight3} &- \sum_{\gamma'\in\Gamma}\sum_{b \in D} A(z^{(a)},\gamma' z^{(b)}) \psi(\gamma \gamma' z^{(b)})=E \psi(\gamma z^{(a)}).
\end{align} 
Crucially, although written as an infinite sum over elements of $\Gamma$, only $q$ terms on the left-hand side of this equation are nonzero, with $A(z^{(a)},\gamma'z^{(b)})=1$. The nonvanishing contributions correspond to those $\gamma'\in\Gamma$ that yield a neighboring site of $z_i$. It is straightforward to determine these $q$ group elements for each $z^{(a)}$, as they must satisfy $d(\gamma\gamma'z^{(b)},\gamma z^{(a)})=d(\gamma'z^{(b)},z^{(a)})=d_0$, with nearest-neighbor distance $d_0$. Here we used the invariance of the hyperbolic distance under isometries.

The Hamiltonian for the infinite lattice is invariant under a simultaneous Fuchsian translation of all sites, i.e. $z_i \to \gamma z_i$ with $\gamma\in \Gamma$. This implies that the choice of the reference unit cell, i.e. the value of $\gamma$, cannot affect the solution of Eq. (\ref{tight3}). Formally, define the group action $\Gamma$ on the Hilbert space of wave functions as $(T_\gamma \psi)(z) = \psi(\gamma^{-1} z)$. Then an overall factor $T_{\gamma^{-1}}$ can be extracted from Eq. (\ref{tight3}) and we arrive at
\begin{align}
 \label{tight4} &- \sum_{\gamma'\in\Gamma}\sum_{b \in D} A(z^{(a)},\gamma' z^{(b)}) \psi(\gamma' z^{(b)})=E \psi(z^{(a)}),
\end{align} 
which is equivalent to setting $\gamma=1_{\mathcal{P}}$. Equation (\ref{tight4}) is one of the central results of this work, and the key application of hyperbolic crystallography to tight-binding Hamiltonians. We accomplished to reduce the eigenvalue problem in Eq. (\ref{tight1}), which needs to be solved for the infinite number of graph sites $z_i$, to a set of $N$ coupled equations for the unit cell sites $z^{(a)}$. Each of these equations features only a finite number of nonvanishing terms. This reduction needs to be compared to the significance of Euclidean crystallography in studying Euclidean lattice models, where the band structure is typically obtained from a few lines of calculation after the unit cell has been identified.

As with every symmetry in quantum mechanics, the "translation invariance" of the Hamiltonian implies that its eigenfunctions belong to irreducible representations (irrep) of $\Gamma$. Assume that these irreps are labeled by $k$, and that $\psi_k$ is a wave function in the corresponding eigenspace of dimension $d(k)\leq \infty$ over $\mathbb{C}$ with energy $E_k$. If $\phi_{km}(z)$, $m=1,\dots,d(k)$, is a basis of the eigenspace, then we write 
\begin{align}
 \label{tight5} \psi_k(z) = \sum_{m=1}^{d(k)}c_{m}(z) \phi_{km}(z),
\end{align}
and the coefficients $c_m$ transform linearly under $\Gamma$ as $\gamma: c_m \mapsto \sum_{m'} D_{mm'}(\gamma) c_{m'}$ with a matrix $D(\gamma)$ satisfying $D(\gamma_1\gamma_2)=D(\gamma_1)D(\gamma_2)$. Writing $c_m^{(a)}=c_m(z^{(a)})$ we arrive at
\begin{align}
 \label{tight6} - \sum_{\gamma'\in\Gamma}\sum_{b \in D} \sum_{m'=1}^{d(k)} A(z^{(a)},\gamma' z^{(b)}) D_{mm'}(\gamma')c_{m'}^{(b)}=E c_m^{(a)}.
\end{align} 
The corresponding number of linear coupled equations for the coefficients $c_m^{(a)}$ that determine the energy $E_k$ is $N\times d(k)$. Together, equations (\ref{tight4}) and (\ref{tight6}) constitute the first step towards computing the eigenvalues and band structure of the tight-binding Hamiltonian $\hat{H}$.

\subsection{Bloch wave theory}\label{SecBloch}

In this section, we sketch the implications of Eq. (\ref{tight6}) for one-dimensional representations ($d(k)=1$). We find that in this case $k \to \textbf{k}=(k_1,\dots,k_{2g})^T$. The corresponding eigenfunctions $\psi_{\textbf{k}}(z)$ are Bloch waves and lead to an intriguing band structure. A detailed study of the related Bloch wave theory for hyperbolic lattices will be presented elsewhere.

For Euclidean lattices, the translation group $\Gamma_{g=1}\simeq \mathbb{Z}^2$ is Abelian and so all irreducible representations are one-dimensional (Bloch's theorem). We label Euclidean translations by $\textbf{n}\in \mathbb{Z}^2$ and have
\begin{align}
 \label{app6} \psi_{\textbf{k}}^{(\rm Eucl)}(\gamma_{\textbf{n}}\textbf{x}) = e^{\rmi \textbf{k}\cdot\textbf{n}} \psi_{\textbf{k}}^{(\rm Eucl)}(\textbf{x})
\end{align}
with the crystal momentum $\textbf{k}$ labeling the irreducible representations. For hyperbolic lattices, $\Gamma_g$ is non-Abelian and so not all irreducible representations are one-dimensional. On the other hand, it is natural to expect that \emph{some} eigenfunctions $\psi(z)$ of $\hat{H}$ transform according to a one-dimensional representation, i.e. satisfy
\begin{align}
 \label{app7} \psi_{\textbf{k}}(\gamma_\mu z) = e^{\rmi k_{\mu}}\psi_{\textbf{k}}(z),
\end{align}
with generalized crystal momentum $\textbf{k}=(k_1,\dots,k_{p_{\rm B}/2})$ and the index $\mu=1,\dots,p_{\rm B}/2$ counting the number of momentum components. We refer to functions $\psi_{\textbf{k}}(z)$ satisfying Eq. (\ref{app7}) as \emph{Bloch waves}. They are also called automorphic forms with respect to the group $\Gamma_g\subset \mathcal{P}$. Note that the condition $X_g=1$ from Eq. (\ref{bra23b}) is automatically satisfied for Bloch waves. Importantly, the number of independent momentum components of $\textbf{k}$ is $2g$. Hence, for hyperbolic Bloch waves ($g>1$), the dimension of coordinate and momentum space differ.

The eigenvalue $E_{\textbf{k}}$ of a Bloch wave with momentum $\textbf{k}$ is obtained from Eq. (\ref{tight6}) by inserting $D(\gamma_\mu)=e^{\rm ik_\mu}$. This results in a Schr\"{o}dinger equation that can be written as
\begin{align}
 \label{app7b} -\sum_{b\in D} \bar{A}_{ab}(\textbf{k}) c^{(b)} = E_{\textbf{k}} c^{(a)}.
\end{align}
For every $\textbf{k}$, the possible eigenvalues $E_{\textbf{k}}$ of Bloch waves follow from diagonalizing the $\textbf{k}$-dependent $N\times N$ matrix $\bar{A}(\textbf{k})$.  We call the single-particle Hamiltonian $\hat{H}_{\rm BW}$ constructed from $\bar{A}(\textbf{k})$ in Eq. (\ref{intro2}) the \emph{Bloch wave Hamiltonian}.

As a nontrivial example, we compute the matrix $\bar{A}(\textbf{k})$ for the ten-site unit cell of the $\{10,3\}$ lattice shown in Fig. \ref{FigBands} in the introduction. We have
\begin{widetext}
\begin{align}
\label{app8} \bar{A}(\textbf{k}) = \begin{pmatrix} 0& 1& 0& 0& 0& e^{\rmi k_1} & 0& 0& 0& 1 \\ 1& 0& 1& 0& 0& 0&  e^{\rmi k_2}& 0& 0& 0 \\ 0& 1& 0& 1& 0& 0& 0&  e^{\rmi k_3}& 0& 0 \\ 0& 0& 1& 0& 1& 0& 0& 0&  e^{\rmi k_4}& 0 \\ 0& 0& 0& 1& 0& 1& 0& 0& 0&  e^{\rmi k_5} \\  e^{-\rmi k_1}& 0& 0& 0& 1& 0& 1& 0& 0& 0 \\ 0&  e^{-\rmi k_2}& 0& 0& 0& 1& 0& 1& 0& 0 \\ 0& 0&  e^{-\rmi k_3}& 0& 0& 0& 1& 0& 1& 0 \\ 0& 0& 0& e^{-\rmi k_4}& 0& 0& 0& 1& 0& 1 \\ 1& 0& 0& 0&  e^{-\rmi k_5} & 0& 0& 0& 1& 0\end{pmatrix}.
\end{align}
\end{widetext}
The generators of the $\{10,5\}$ Bravais lattice satisfy $\gamma_1\gamma_2^{-1}\gamma_3\gamma_4^{-1}\gamma_5=1_{\mathcal{P}}$, see Appendix \ref{AppFuchs}, which implies $k_5=-(k_1-k_2+k_3-k_4)$. For any given $\textbf{k}=(k_1,k_2,k_3,k_4)$, it is straightforward to determine the eigenvalues of $\bar{A}(\textbf{k})$. An example band structure is shown in Fig. \ref{FigBands} in the introduction.

\section{Summary and Outlook}

In this work, we have developed a crystallography for infinite hyperbolic $\{p,q\}$ lattices. By utilizing the notion of patterns on Riemann surfaces, we identified regular $\{p_{\rm B},q_{\rm B}\}$ lattices that constitute Bravais lattices. We then explicitly constructed examples of $\{p,q\}$ lattices whose Bravais lattices are of this type and discussed the associated unit cells. Among the examples are five infinite families and a handful of exceptional cases for genus two and three, many of which we expect to be relevant for advancing our understanding of hyperbolic lattices in future studies. To the best of our knowledge, no such list of hyperbolic lattices and their Bravais lattices existed before. The explicit formulas for Fuchsian translation groups constructed in this work bridge the gap between abstract mathematics and concrete calculations, and will be crucial in practical applications. The present work, therefore, lays the foundation for applying powerful concepts of solid state physics, such as crystal momentum or Bloch waves, to hyperbolic lattices.

A number of pressing questions are raised by the results presented here. Here, as an outlook, we point out two of them.

\emph{(1) Classification of $\{p,q\}$ lattices.} In the present work, we only searched for hyperbolic lattices with regular Bravais lattices. We do not know whether every $\{p,q\}$ lattice has a regular Bravais lattice. Given the small set of lattices we identified from our systematic search, we expect that large classes of $\{p,q\}$ lattices have irregular Bravais lattices. For instance, such a Bravais lattice can have a fundamental domain that is a polygon whose internal angles are not all equal. This expectation is also supported by the Euclidean case, where only two out of five Bravais lattices are regular. Furthermore, even within the set of regular Bravais lattices, we only discussed those of type $\{4g,4g\}$ or $\{2(2g+1),2g+1\}$, because their Fuchsian translation group $\Gamma_g$ is easily constructed. On the other hand, more regular Bravais lattices such as $\{4(2g-1),3\}$ and $\{6(2g-1),3\}$ follow from $F_0=1$. We found that some $\{p,q\}$ lattices seem to feature these Bravais lattices, but leave a conclusive study for future investigation.

\emph{(2) Representation theory.} Equation (\ref{tight6}) constitutes the first step towards solving the spectral problem for the tight-binding Hamiltonian on an infinite hyperbolic lattice, i.e.~determining the single-particle energy band structure. We have outlined how one-dimensional representations of the group $\Gamma_g$ lead to Bloch wave theory for hyperbolic lattices. It will be extremely exciting to study higher-dimensional representations and how their appearance is linked to the noncommutativity of spatial isometries. First important results in this direction have been obtained in Ref. \cite{maciejko2021automorphic}.

\acknowledgements \noindent The authors thank C. Baldwin, P. Bienias, R. Belyansky, A. Fritzsche, T. Helbig, T, Hofmann, S. Imhof, T. Kie\ss{}ling, F. Kohr, J. Koll\'{a}r, R. Mazzeo, P. Sarnak, A. Stegmaier, J. Szmigielski, L. Upreti for inspiring discussions. IB acknowledges support from the University of Alberta startup fund UOFAB Startup Boettcher. AVG acknowledges funding by the U.S. Department of Energy Award No. DE-SC0019449, ARO MURI, DoE ASCR Quantum Testbed Pathfinder program (award No. DE-SC0019040), DoE ASCR Accelerated Research in Quantum Computing program (award No. DE-SC0020312), NSF PFCQC program, AFOSR, and AFOSR MURI. AK acknowledges support from a University of Maryland startup fund. JM was supported by Natural Sciences and Engineering Research Council of Canada (NSERC) Discovery Grants Nos. RGPIN-2020-06999 and RGPAS-2020-00064; the Canada Research Chairs (CRC) Program; Canadian Institute for Advanced Research (CIFAR); and a Government of Alberta Major Innovation Fund (MIF) Grant. SR acknowledges support from an NSERC Discovery Grant number RGPIN-2017-04520, the Canada Foundation for Innovation (CFI) John R. Evans Leaders Fund, and the University of Saskatchewan. JM and SR acknowledge support from the Tri-Agency New Frontiers in Research (Exploration) Fund and a Pacific Institute for the Mathematical Sciences (PIMS) Collaborative Research Group grant. RT is funded by the Deutsche Forschungsgemeinschaft (DFG, German Research Foundation) through Project-ID 258499086 - SFB 1170 and through the W\"{u}rzburg-Dresden Cluster of Excellence on Complexity and Topology in Quantum Matter - ct.qmat Project-ID 390858490 - EXC 2147.

\begin{appendix}

\section{Generating finite hyperbolic lattices}\label{SecGen}

In this section, we show how to efficiently and systematically create large hyperbolic lattices by applying the Fuchsian translation group to a single unit cell.

One way of computing the coordinates of a large, regular hyperbolic tessellation of the Poincar\'{e} disk is to start with a single $p$-gon and apply products of the generators $A$ and $B$ of the proper triangle group $\Delta^+(p,q,2)$ in Eq. (\ref{bra13}). This method is conceptually simple and can be applied for any $\{p,q\}$ lattice. However, since elements from $\Delta^+(p,q,2)$ have fixed points, this method suffers the drawback that lattice sites are duplicated several times with every iteration. Therefore, at the end of the procedure, the duplicated sites need to be identified and eliminated, which can be numerically challenging as the sites accumulate close to the Poincar\'{e} disk boundary for large lattices. In addition, creating the lattice this way does not immediately yield a systematic labeling of sites.

\begin{figure*}[t!]
\centering
\includegraphics[width=18cm]{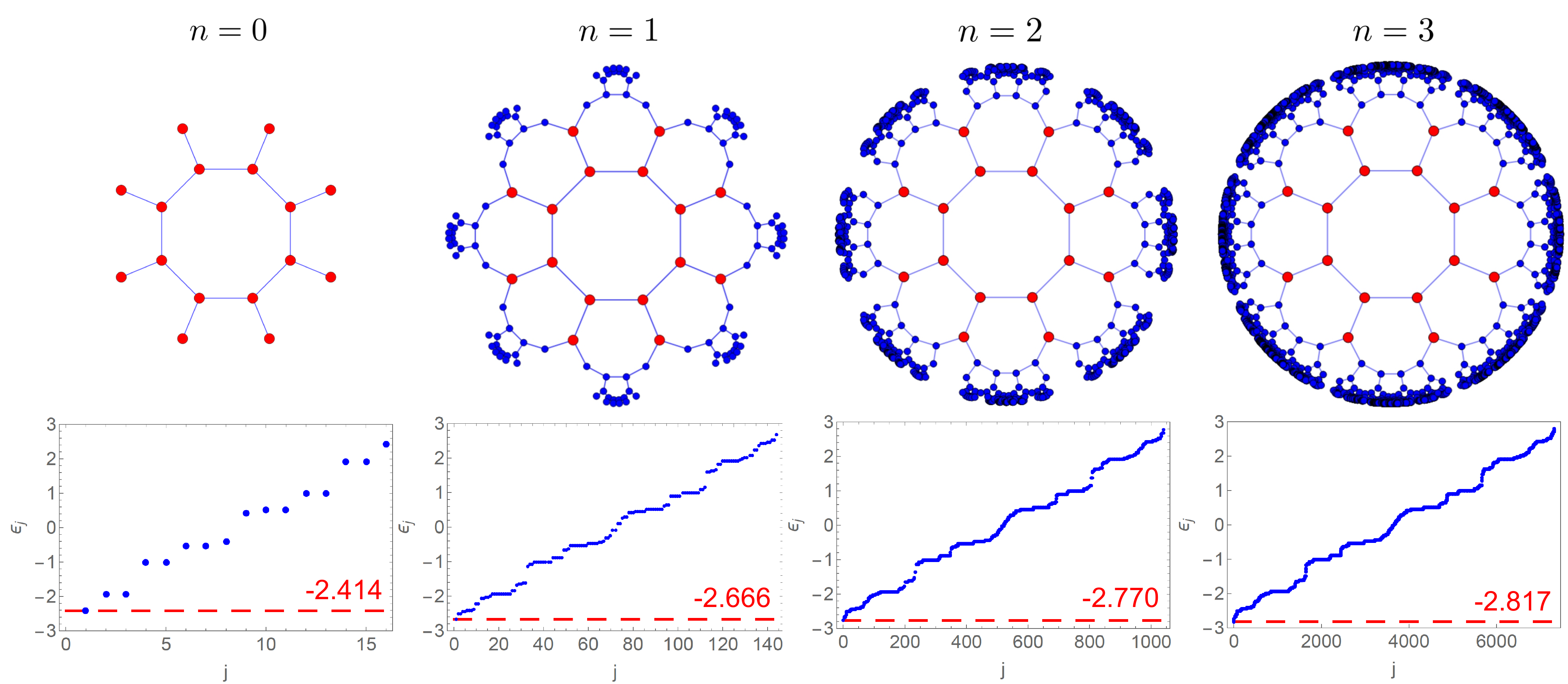}
\caption{We show how to efficiently create large samples of the $\{8,3\}$ lattice, without repetitions of sites, by applying the Fuchsian translation group $\Gamma_{2}$ to the 16-site unit cell. The $n^{\rm th}$ generation consists of all points that are obtained by applying words in the generators of $\Gamma_2$ of length up to $n\geq0$. The corresponding number of sites in the $n^{\rm th}$ generation is 16 ($n=0$), 144 ($n=1$), 1040 ($n=2$), and 7312 ($n=3$). The plots at the bottom show the corresponding eigenvalues $\vare_j$ of the adjacency matrix. The spectrum of the latter is contained in the interval $(-3,3)$. The lowest eigenvalue (highlighted red in each plot) has a sizable gap from $-3$, a characteristic feature of hyperbolic lattices.}
\label{FigGrow83}
\end{figure*}

An alternative and efficient way of computing the coordinates of large hyperbolic lattices is implied by the results presented here. For this, we first need to digress to discuss the nature of elements of the Fuchsian translation group $\Gamma_g$. Every element of $\Gamma_g$ is a product of the generators $\gamma_\mu,\ \mu=1,\dots,p_{\rm B}/2$, and their inverses. Since it would be cumbersome to always mention the inverses separately, we utilize that $(\gamma_\mu)^{-1} = \gamma_{p_{\rm B}/2+\mu}$ and will hereafter refer to $\gamma_\mu$ with $\mu=1,\dots,p_{\rm B}$ as the generators.

Every $\gamma\in \Gamma_g$ is then a word of specific length $n$ in the generators $\gamma_\mu$, i.e. a product of $n$ generators with a well-defined order. For a word of length $n$ we write
\begin{align}
 \label{word1} \gamma=  \gamma_{\mu_1}\cdots \gamma_{\mu_n}
\end{align}
with $\mu_i \in\{1,\dots,p_{\rm B}\}$. The $n$-tuple or vector $\vec{\mu}=(\mu_1,\dots,\mu_n)$ specifies the element $\gamma$. Thus the discrete, but infinite set of nontrivial Fuchsian translations can be labeled by integer vectors $\vec{\mu}$ of arbitrary length $n\geq 1$, and the discrete index $i$ in Eq. (\ref{app1}) corresponds to a discrete index $(a,\vec{\mu})$.

To determine the number of words of length $n$ within a group that is generated by a finite number of generators is called the \emph{word problem} for the group. The number of naive products of length $n$ is $p_{\rm B}^n$ in our case, but the number of words of length $n$ is smaller. First, whenever the combination $\gamma_\mu \gamma_\mu^{-1}=1$ appears, the word size is reduced. Second, constraints like $X_g=1$ in $\Gamma_g$ lead to further reductions. As an example, for the $\{8,8\}$ Bravais lattice with eight generators, we have $\gamma_1\gamma_2^{-1}\gamma_3\gamma_4^{-1} = \gamma_4^{-1}\gamma_3\gamma_2^{-1}\gamma_1$ and other resulting relations, and the number of words of length $n=1,\ 2,\ 3$ is $8,\ 56,\ 392$, whereas the number of naive products is $8^n=8,\ 64,\ 512$. In practice, solving the word problem is not difficult when using the particular representation of the generators $\gamma_\mu$ from Eq. (\ref{bra22}).

Let us now describe our algorithm to create hyperbolic lattices. Assume the $\{p,q\}$ lattice has a unit cell of size $N$ with coordinates
$\{z^{(1)},\dots, z^{(N)}\}\subset \mathbb{D}$ and a regular $\{p_{\rm B},q_{\rm B}\}$ Bravais lattice. We define the $n^{\rm th}$ generation lattice as the set of all points that are generated by applying words of length up to $n$ in the generators of $\Gamma_{\{p_{\rm B},q_{\rm B}\}}$ to the unit cell. The $0^{\rm th}$ generation is just the unit cell, the $1^{\rm st}$ generation contains the unit cell and and all points obtained from applying each generators once, hence $(p_{\rm B}+1)N$ total sites, and so on. In Fig. \ref{FigGrow83} we show, as an example, how the $\{8,3\}$ lattice, with 16-site unit cell and $\{8,8\}$ Bravais lattice, is generated in this manner. The lowest eigenvalue of the adjacency matrix is found to be larger than $-3$ (with coordination number $q=3$ in the example) for all $n$. This gap is a characteristic feature of hyperbolic lattices and surfaces and is expected to converge to a nonzero value for large $n$ \cite{paschke,Higuchi,kollar2019line,PhysRevA.102.032208}.

Let us comment on a fine point regarding the number of independent generators. Clearly, the group $\Gamma_g$ is independent of $p_{\rm B}$ and always has $4g$ independent generators. Nevertheless, when generating a $\{p,q\}$ lattice whose Bravais lattice is the $\{2(2g+1),2g+1\}$ lattice, one can choose to either work with the $4g$ independent generators or to use all $4g+2$ generators. What changes is the number of words of length $n$ that can be composed from these generators, and hence the number of sites in the $n^{\rm th}$ generation lattice, but the procedure is not afflicted otherwise. Using all $4g+2$ generators has the advantage of obtaining a radially symmetric lattices in each generation, which may be favorable in applications.

\section{Fuchsian translation groups}\label{AppFuchs}

The constraint $X_{\{p_{\rm B},q_{\rm B}\}}=1$ in Eq. (\ref{bra23}) generalizes the Euclidean cases from Eqs. (\ref{bra6}) and (\ref{bra10}) in an interesting manner to higher genera. We first consider $\{4g,4g\}$ Bravais lattices with $g\geq 2$. In this case, the generator $\gamma_1$ has the simple form
\begin{align}
  \label{fuchs1} \gamma_1^{\{4g,4g\}}= \frac{1}{\sqrt{1-r_{0,\rm B}^2}}\begin{pmatrix} \sqrt{1+r_{0,\rm B}^2} & \sqrt{2} r_{0,\rm B} \\ \sqrt{2} r_{0,\rm B} & \sqrt{1+r_{0,\rm B}^2} \end{pmatrix},
\end{align}
with $r_{0,\rm B}=\sqrt{\cos(\alpha_{\rm B})}$. The remaining $\gamma_\mu$ follow from Eq. (\ref{bra22}). For $g=2$, we have
\begin{align}
 \label{fuchs2} X_{\{8,8\}} = \gamma_1\gamma_2^{-1}\gamma_3 \gamma_4^{-1} \gamma_1^{-1}\gamma_2\gamma_3^{-1}\gamma_4,
\end{align}
which generalizes to 
\begin{align}
 \label{fuchs3} X_{\{4g,4g\}} = \gamma_1\gamma_2^{-1} \cdots \gamma_{2g-1}\gamma_{2g}^{-1} \gamma_1^{-1}\gamma_2\cdots \gamma_{2g-1}^{-1}\gamma_{2g}
\end{align}
for any $g\geq 2$.

For the $\{2(2g+1),2g+1\}$ Bravais lattices with $g\geq 2$ we have
\begin{align}
 \label{fuchs4} \gamma_1 = \frac{1}{\sqrt{1-r_{0,\rm B}^2}}\begin{pmatrix} 1+r_{0,\rm B}^2 & r_{0,\rm B}\sqrt{3+r_{0,\rm B}^2} \\ r_{0,\rm B}\sqrt{3+r_{0,\rm B}^2} &  1+r_{0,\rm B}^2 \end{pmatrix}.
\end{align}
Since the fundamental polygon has $p_{\rm B}=4g+2$ sides, there are, naively, two more generators than for the $\{4g,4g\}$ Bravais lattice. However, both lattices tessellate surfaces of genus $g$ and the number of independent generators should be equal. The issue is resolved, as in the case of the hexagonal $\{6,3\}$ lattice, by the fact that the translation $\gamma_{2g+1}$ is not independent of the remaining $\gamma_1,\dots,\gamma_{2g}$. We have
\begin{align}
 \label{fuchs5} \gamma_1 \gamma_2^{-1}\cdots \gamma_{2g}^{-1} \gamma_{2g+1} =(-1)^{g+1}\mathbb{1}.
\end{align}
The remaining $2g$ independent generators satisfy the same constraint as for the $\{4g,4g\}$ lattice, i.e.~we have
\begin{align}
 \label{fuchs6} X_{\{2(2g+1),2g+1\}} = X_{\{4g,4g\}},
\end{align}
and, therefore,
\begin{align}
 \label{fuchs7} \Gamma_{\{2(2g+1),2g+1\}} = \Gamma_{\{4g,4g\}}.
\end{align}

\section{Sublattice structure}\label{AppSub}

\begin{figure}[t!]
\centering
\includegraphics[width=6cm]{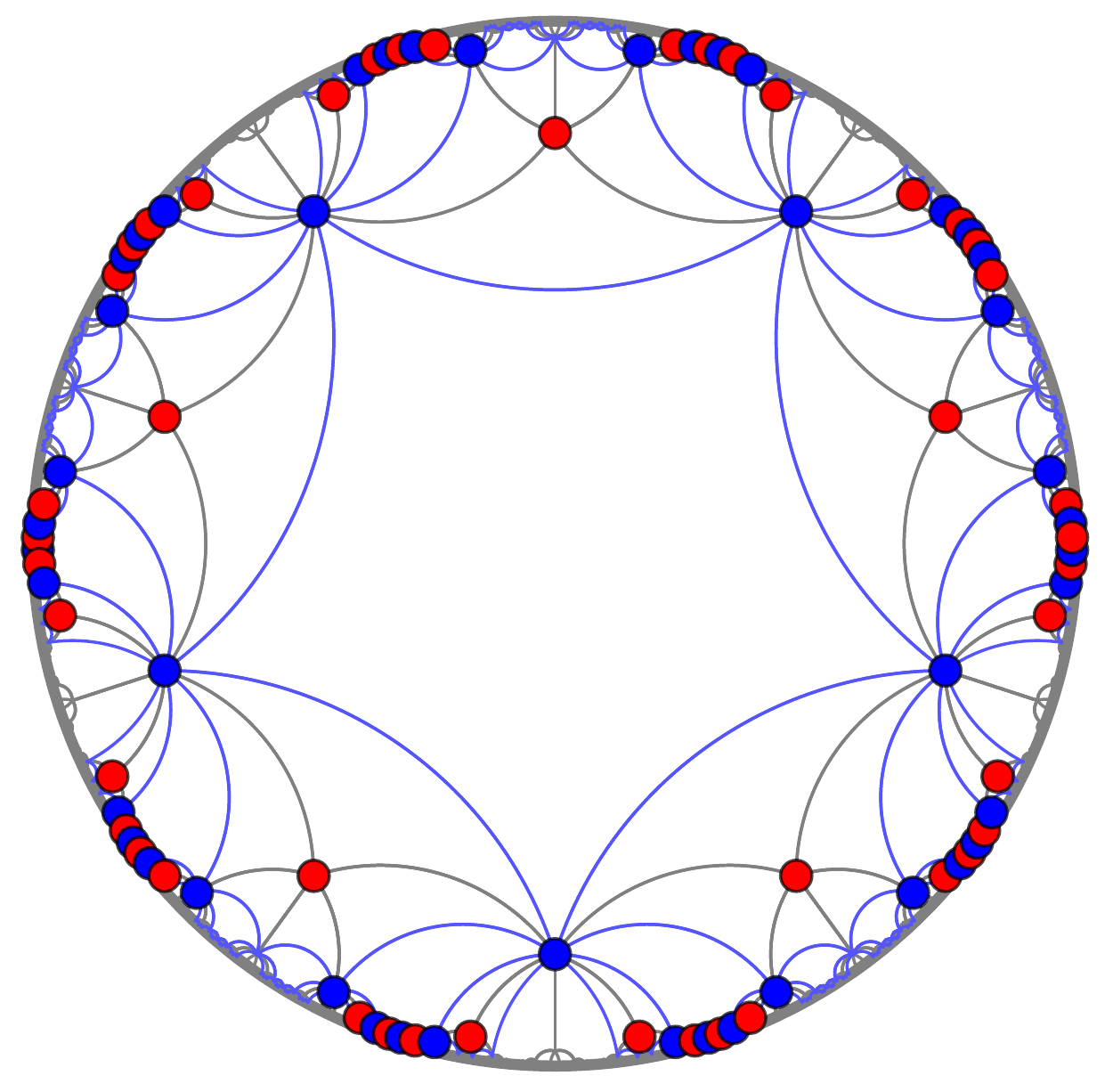}
\caption{The bipartite $\{2(2g+1),2g+1\}$ lattice can be divided into two $\{2g+1,2(2g+1)\}$ lattices. This well-known fact from Euclidean lattices, where the $\{3,6\}$ triangular lattice is the sublattice of the $\{6,3\}$ honeycomb lattice, generalizes to hyperbolic lattices of higher genus. Here we show the case of $g=2$, with the $\{10,5\}$ lattice indicated by the gray geodesics, and a selection of sites of the two sublattices marked red and blue. One of the $\{5,10\}$ sublattices is indicated by blue geodesics.}
\label{Fig105Sub}
\end{figure}

An interesting analogy to the Euclidean case can be observed in the third family in Table \ref{TabUnits2}. Placing a single site ($N=1$) on a vertex of the fundamental polygon of the $\{2(2g+1),2g+1\}$ lattice, we generate the $\{2g+1,2(2g+1)\}$ lattice instead. This implies that the sites of the $\{2g+1,2(2g+1)\}$ lattice form the sublattice of the bipartite $\{2(2g+1),2g+1\}$ lattice. In the Euclidean case, we obtain the well-known fact that the $\{3,6\}$ triangular lattice is the sublattice of the $\{6,3\}$ honeycomb lattice. However, a subtle difference arises in the hyperbolic case. While the $\{3,6\}$ lattice coincides with the next-to-nearest neighbor graph of the $\{6,3\}$ lattice, which we define here by connecting any two sites of the $\{6,3\}$ lattice that are separated by two adjacent edges, this is not true for $g>1$, because not all sites that are separated by two adjacent edges have the same hyperbolic distance. We visualize the hyperbolic case for $g=2$ in Fig. \ref{Fig105Sub}.

\section{Distance spectrum}\label{AppDist}

In order to decide whether a given unit cell $D=\{z^{(1)},\dots,z^{(N)}\}$ and regular $\{p_{\rm B},q_{\rm B}\}$ Bravais lattice generate the infinite $\{p,q\}$ lattice, denoted $\Lambda_{\{p,q\}}$, we have to show that
\begin{align}
 \label{distA} \Gamma_{\{p_{\rm B},q_{\rm B}\}} D \stackrel{!}{=} \Lambda_{\{p,q\}},
\end{align}
where $\Gamma_{\{p_{\rm B},q_{\rm B}\}}$ is the Fuchsian translation group of the Bravais lattice. In the remainder of this section we write
\begin{align}
 \Lambda := \Lambda_{\{p,q\}},\ \Gamma:=\Gamma_{\{p_{\rm B},q_{\rm B}\}},\ \Lambda':= \Gamma D,
\end{align}
and Eq. (\ref{distA}) becomes $\Lambda'\stackrel{!}{=}\Lambda$.

Verifying Eq. (\ref{distA}) may sound simple, as one merely needs to check that every $z_i \in \Lambda'$ is contained in $\Lambda$, and vice versa. However, in practice, no parametrization of the coordinates of the full hyperbolic lattice $\Lambda$ exists, rather only powerful algorithms to construct a finite subset of it. Furthermore, it is impossible in practice to compute $\Lambda'$, which is an infinite set and would require applying the generators infinitely often. Rather, one applies finite-length words in the generators $\gamma_\mu$ to $D$. Due to the exponential proliferation of words, typical feasible word lengths range from $2$ to $5$, although this number depends on $p$ and $q$. So, in practice, we can only access a finite subset of either side of Eq. (\ref{distA}), and neither of these subsets is strictly contained in the other.

To show that Eq. (\ref{distA}) is true, we need to confirm that
\begin{itemize}
 \item[(1)] $\Lambda'$ does not contain additional sites that are not in $\Lambda$. (That is $\Lambda'\subset \Lambda$)
 \item[(2)] No sites in $\Lambda'$ are doubled.
 \item[(3)] No sites of $\Lambda$ are left out in $\Lambda'$: This means that, for any site $z_i\in\Lambda$, we can make $\Lambda'$ large enough so that $z_i\in\Lambda'$. (That is $\Lambda\subset \Lambda'$.)
\end{itemize}
We imply here that we can only compute large, but finite samples $\Lambda'$. By making these samples large enough and verifying (1) and (3), we obtain reliable evidence that Eq. (\ref{distA}) is true. Note that condition (2) is additional; it shows that $\Gamma$ is a normal subgroup and that the unit cell $D$ has been identified correctly, i.e. the split in Eq. (\ref{bra1}) is unique.

The practical solution to verifying (1) and (2) consists in identifying the \emph{distance spectrum} as a unique fingerprint of any hyperbolic $\{p,q\}$ lattice. Given the hypothetical infinite lattice $\Lambda$, we compute the list of values $d(z_i,z_j)$, where $i,j$ runs over all distinct lattice sites. Crucially, only a certain discrete set of numbers appears in this list, which is fully specified by $p$ and $q$. We call this list the distance spectrum. The first entry is given by the nearest-neighbor distance
\begin{align}
 \label{dist1} d_0 = d(r_0,r_0 e^{2\pi\rmi/p}).
\end{align} 
Some examples are shown in Table \ref{TabDist}. Given a sufficiently large finite subset of either $\Lambda$ or $\Lambda'$, we can compute their truncated distance spectra. (The finite subset needs to be reasonable, i.e.~contain at least one pair of nearest neighbors, next-nearest neighbors, etc.) We call the distance spectrum obtained from the finite sample of $\Lambda$ the reference spectrum, and ignore its difference from the full (infinite) distance spectrum $\delta_{\{p,q\}}$ in the following.

\renewcommand{\arraystretch}{1.3}
\begin{table}[t!]
\begin{center}
\begin{tabular}{|c||c|c|c|}
\hline \  \  & \ $\{7,3\}$ \ & \ $\{8,3\}$ \ & \ $\{9,3\}$ \ \\
\hline\hline \ $d_0/(2\kappa)$ \  & \ $0.283128 $ \ & \ $0.363520 $ \ & \ $0.409595 $ \ \\
\hline \ $d_1/(2\kappa)$ \  & \ $0.496385 $ \ & \ $0.641645 $ \ & \ $0.409596 $ \ \\
\hline \ $d_2/(2\kappa)$ \  & \ $0.606789 $ \ & \ $0.806689 $ \ & \ $0.726012 $ \ \\
\hline \ $d_3/(2\kappa)$ \  & \ $0.753167 $ \ & \ $0.860706 $ \ & \ $0.927539 $ \ \\
\hline \ $d_4/(2\kappa)$ \  & \ $0.887104 $ \ & \ $0.970155 $ \ & \ $1.02404 $ \ \\
\hline
\end{tabular} 
\end{center}
\label{TabDist}
\caption{Distance spectra for the $\{7,3\}$, $\{8,3\}$, and $\{9,3\}$ lattices. We display the first low-lying five entries. The distance spectrum acts as a unique fingerprint of the hyperbolic lattice and can be used to decide whether a set of points $\Lambda'=\{z_i\}$ could be a subset of the infinite $\{p,q\}$-lattice. The first entry, $d_0$, is the hyperbolic distance between nearest neighbors.}
\end{table}
\renewcommand{\arraystretch}{1}

To verify (1), we then compute the distance spectrum of $\Lambda'$. It probes the local surrounding of every single site $z_i\in \Lambda'$. If there was any site $\tilde{z}_i\in \Lambda'$ that is not part of $\Lambda$, then the numbers $d(\tilde{z}_i,z_j)$ appearing in the distance spectrum of $\Lambda'$ would not be contained in $\delta_{\{p,q\}}$. In practice, we find that this test is very sensitive to detecting lattice sites that should not be there, with deviations showing up in the first few entries of the distance spectra. In contrast, if two lattices seem to match, i.e. $\Lambda=\Lambda'$, the first deviating entries in their truncated distance spectra of large samples show up at the hundredths or thousandths position, giving strong evidence that (1) is true. As a byproduct, by computing the distance spectrum of $\Lambda'$, we can show that no sites are doubled, i.e. (2) is true, by verifying that there is no entry $d(z_i,z_j)=0$ in the distance spectrum. (The latter is computed for $i\neq j$, and so should not contain zeros.)

The efficiency of the method of comparing distance spectra stems from the fact that the agreement of the distance spectra of $\Lambda'$ and $\Lambda$ is a necessary condition and, therefore, incompatible (or wrong) Bravais lattices yield a negative result and can be excluded even for small sample lattices. Let us also point out that one can decide whether a given $\{p_{\rm B},q_{\rm B}\}$ lattice can be the Bravais lattice of a given $\{p,q\}$ lattice without the precise knowledge of the unit cell $D$. Since the Fuchsian translation group $\Gamma$ is a subgroup of the symmetry group of the $\{p,q\}$ lattice, we have
\begin{align}
 \label{distB} \Gamma S \subset \Lambda
\end{align}
for \emph{every} subset $S\subset \Lambda$. By choosing a sufficiently large $S$ (such that $D\subset S$), we obtain equality in Eq. (\ref{distB}). However, $S\neq D$ violates (2) and so will create zero-entries in the distance spectrum of $\Gamma S$.

Let us now comment on condition (3), which is less straightforward to check. Assuming that (1) and (2) hold, the validity of (3) ensures that the unit cell has enough elements to generate the whole lattice $\Lambda$. Indeed, if we removed a few sites $z^{(a)}$ from $D$, conditions (1) and (2) would still be satisfied, but, nonetheless, $\Lambda \neq \Lambda'$. This implies a method how to test (3): Assume our choice of $D=\{z^{(1)},\dots,z^{(N)}\}$ satisfies (1) and (2). Condition (3) can only be violated if there is a lattice site $z^{(N+1)}\notin D$ such that the union $D \cup \{z^{(N+1)}\}$ still satisfies (2). Reasonable choices for $z^{(N+1)}$ are limited in practice. In all cases we identified, the unit cell is either made from a connected graph, i.e.~every $z^{(a)}$ in $D$ is nearest neighbor to at least one $z^{(b)}$ in $D$, or the sites from $D$ are taken from the central $p$-gon of the $\{p,q\}$ lattice. Thus we only need to consider the finite set
\begin{align}
 \nonumber D' ={}& \{ z_i\in\Lambda:\ d(z_i,z^{(a)})=d_0\ \text{for some }z^{(a)}\in D\} \\
 \label{distC} &\cup \{ r_0 e^{2\pi \rmi n/p}e^{\rmi \chi},\ n=1,\dots,p\}.
\end{align}
The phase $e^{\rmi \chi}$ is for adjusting the overall rotation of the lattice within the unit cell and is easily found in each case. If condition (2) fails for the enlarged unit cell $D\cup \{z^{(N+1)}\}$ for every $z^{(N+1)}\in D'$, then $D$ is big enough to generate the whole lattice $\Lambda$.
(Since $z^{(N+1)}\in\Lambda$, it is trivial that $\Gamma z^{(N+1)}$ satisfies (1).) We carried out this test of condition (3) for all lattices discussed here and found that no sites are missing in the listed unit cells.

An alternative way to test for condition (3) is to use a finite set of elements from $\Gamma$ to generate, starting from $D$, one $p$-gon and all its neighboring $p$-gons of the $\Lambda$-lattice. This implies, iteratively, that all polygons of $\Lambda$ can be generated by applying elements from $\Gamma$. This method can be applied to almost all cases, except, for instance, the $\{7,3\}$ lattice with a very big unit cell. Note that the condition $X=1_{\mathcal{P}}$ ensures that, after applying sufficiently many generators from $\Gamma$, we eventually obtain closed $p$-gons.

\section{Systematic search for unit cells and Bravais lattices}\label{AppSearch}

Assume we are given a $\{p,q\}$ lattice $\Lambda_{\{p,q\}}$ and want to determine its unit cell and Bravais lattice. Assume further that the Bravais lattice is a regular $\{p_{\rm B},q_{\rm B}\}$ lattice. A number of necessary conditions are implied by this, which can be used for a systematic search of hyperbolic lattices and their regular Bravais lattices. For this, note that every generator $\gamma_\mu$ of the Bravais lattice maps a center of a Bravais lattice face to a center of a Bravais lattice face. Put differently, $\gamma_\mu$ maps a vertex of the (dual) $\{q_{\rm B},p_{\rm B}\}$ lattice $\Lambda_{\{q_{\rm B},p_{\rm B}\}}$ to a vertex in $\Lambda_{\{q_{\rm B},p_{\rm B}\}}$. Since the Bravais lattice is rotation symmetric about the centers of its faces, it is natural to expect that each center of a Bravais lattice face  is also the center of some face of the $\{p,q\}$ lattice. Consequently, every site of the $\{q_{\rm B},p_{\rm B}\}$ lattice is also a site of the (dual) $\{q,p\}$ lattice. We thus formulate the \emph{dual lattice criterion} that, if $\{p_{\rm B},q_{\rm B}\}$ is the Bravais lattice of $\{p,q\}$ under the above conditions, then
\begin{align}
 \label{dual1} \Lambda_{\{q_{\rm B},p_{\rm B}\}} \subset \Lambda_{\{q,p\}}.
\end{align}
Although just a necessary condition to match partners $\{p,q\}$ and $\{p_{\rm B},q_{\rm B}\}$, it yields a very efficient and selective search algorithm.

A first consequence of Eq. (\ref{dual1}) is that the distance spectrum of $\{q_{\rm B},p_{\rm B}\}$ is contained in the distance spectrum of $\{q,p\}$. In particular, this is true for the hyperbolic nearest-neighbor distance in the $\{q_{\rm B},p_{\rm B}\}$ lattice given by
\begin{align}
 \label{dual2} d_{0\{q_{\rm B},p_{\rm B}\}} = d(r_{0\rm B},r_{0\rm B}e^{2\pi \rmi/q_{\rm B}}).
\end{align}
Hence we arrive at the necessary condition
\begin{align}
 \label{dual3} \ d_{0\{q_{\rm B},p_{\rm B}\}} \in \delta_{\{q,p\}}.
\end{align}
For a given $(p_{\rm B},q_{\rm B})$, it is numerically straightforward to identify all potential solutions $(p,q)$ from Eq. (\ref{dual3}), as for sufficiently large $p$ and $q$, the smallest entry $d_{0\{q,p\}}$ on the right-hand side becomes too large for the inclusion to be valid.

A second consequence of Eq. (\ref{dual1}) follows from the fact that the $\{p,q\}$ lattice is left invariant under rotations by $2\pi/p$ around the centers of its faces. If such a rotation point is also the center of a face of the Bravais lattice, then we expect it to leave the Bravais lattice invariant. Hence it must be a rotation by an integer multiple of $2\pi/p_{\rm B}$. We conclude that
\begin{align}
 \label{dual4} \frac{2\pi}{p} = n \frac{2\pi}{p_{\rm B}}
\end{align} 
or
\begin{align}
 \label{dual5}  p_{\rm B} = n \cdot p
\end{align}
with an integer $n\geq 1$.

Taken together, conditions (\ref{dual3}) and (\ref{dual5}) yield a small number of possible candidate $\{p,q\}$ lattices for a given Bravais lattice $\{p_{\rm B},q_{\rm B}\}$. The possible values for $p$ are bounded from above by $p\leq p_{\rm B}$, and the maximal values for $q$ that need to be considered are effectively limited from above by Eq. (\ref{dual3}). After the candidate values for $(p,q)$ have been found, we apply the distance spectrum method described in App. \ref{AppDist} to probe if the $\{p_{\rm B},q_{\rm B}\}$ lattice is truly the Bravais lattice of the $\{p,q\}$ lattice. This involves finding the correct unit cell, which is restricted by a combination of symmetry considerations and the value of $V_0$. All the examples in Tables \ref{TabUnits2} and \ref{TabUnits} have been obtained with this search algorithm.

\end{appendix}

\bibliography{refs_crystal}

\end{document}